\documentclass[aps,pre,eprint,groupedaddress]{revtex4-2}

\usepackage{graphicx}

\begin{document}

\title{The $q$-neighbor Ising model on multiplex networks with partial overlap of nodes}

\author{A.\ Krawiecki and T.\ Gradowski}       

\affiliation{Faculty of Physics,
Warsaw University of Technology, \\
Koszykowa 75, PL-00-662 Warsaw, Poland}

\begin{abstract}
The $q$-neighbor Ising model for the opinion formation on multiplex networks with two layers in the form of random graphs (duplex networks), the partial overlap of nodes, and LOCAL\&AND spin update rule was investigated by means of the pair approximation and approximate Master equations as well as Monte Carlo simulations. Both analytic and numerical results show that for different fixed sizes of the $q$-neighborhood and finite mean degrees of nodes within the layers the model exhibits qualitatively similar critical behavior as the analogous model on multiplex networks with layers in the form of complete graphs. However, as the mean degree of nodes is decreased the discontinuous ferromagnetic transition, the tricritical point separating it from the continuous transition and the possible coexistence of the paramagnetic and ferromagnetic phases at zero temperature occur for smaller relative sizes of the overlap. Predictions of the simple homogeneous pair approximation concerning the critical behavior of the model under study show good qualitative agreement with numerical results; predictions based on the approximate Master equations are usually quantitatively more accurate, but yet not exact. Two versions of the heterogeneous pair approximation are also derived for the model under study, which, surprisingly, yield predictions only marginally different or even identical to those of the simple homogeneous pair approximation. In general, predictions of all approximations show better agreement with the results of Monte Carlo simulations in the case of continuous than discontinuous ferromagnetic transition.
\end{abstract}

% multiplex networks; phase transitions; pair approximation; $q$-voter model.

\maketitle

%           ---
%            | 
%            |
%            |
%           ---

\section{Introduction}
\label{intro}

Investigation of the opinion formation process by means of nonequilibrium models has become a firmly established research field in statistical physics in the last decades \cite{Castellano09a}. Many results in this area were obtained using models with agents' opinions represented by spins with discrete (in most cases two) states obeying stochastic dynamics described by various rates at which agents change (e.g., flip) their opinions, e.g., the majority-vote model \cite{Oliveira92, Chen15, Chen17, Nowak20, Chen20, Kim21}, the noisy voter model \cite{Granovsky95, Carro16, Peralta18a}, different versions of the noisy nonlinear and $q$-voter model \cite{Castellano09, Nyczka12, Moretti13, Jedrzejewski17, Jedrzejewski22, Peralta18, Vieira18, Vieira20, Abramiuk21, Nowak21} and the $q$-neighbor Ising model \cite{Jedrzejewski15, Park17, Jedrzejewski17a, Chmiel18}. In particular, much effort was devoted to determining conditions under which the above-mentioned models exhibit phase transition from a disordered paramagnetic (PM) state in which each opinion appears with the same probability to an ordered ferromagnetic (FM) state with one dominant opinion as the parameter controlling the level of stochastic noise in the model is varied, measuring the agents' uncertainty in decision making. In this context the presence of the first-order FM transition, or even transition to a frozen FM phase is of prime importance, with abrupt occurrence of a dominant opinion as well as possible hysteresis and bistability of the PM and FM phases \cite{Chen17,Nowak20,Nyczka12,Moretti13,Jedrzejewski17,Jedrzejewski22,Peralta18,Vieira18,Vieira20,Abramiuk21,Nowak21,Jedrzejewski15,Chmiel18}. Following the growing interest in the dynamical processes on complex networks \cite{Dorogovtsev08} agents in the models for the opinion formation are often located in the nodes and interact via edges of complex networks reflecting a complicated structure of social interactions \cite{Chen15,Chen17,Nowak20,Chen20,Kim21,Carro16,Moretti13,Jedrzejewski17,Jedrzejewski22,Peralta18,Vieira20,Abramiuk21,Nowak21,Jedrzejewski17a,Chmiel18}. In this case analytic predictions concerning the critical behavior of the models based on the mean-field approximation (MFA) need not exhibit quantitative agreement with results of Monte Carlo (MC) simulations, hence, more accurate approaches based on the pair approximation (PA) \cite{Vazquez08,Pugliese09,Gleeson11,Gleeson13,Peralta20} and approximate Master equations (AMEs) \cite{Gleeson11,Gleeson13,Peralta20} were applied to describe theoretically the observed phase transitions \cite{Peralta18a,Jedrzejewski17,Jedrzejewski22,Peralta18,Vieira20,Abramiuk21,Chmiel18}.

Recently much attention has been devoted to combining complex networks in order to create even more complicated and heterogeneous structures known in general as "networks of networks" \cite{Buldyrev10,Gao11,Gao12,Boccaletti14,Gao22}. An important class of such structures is formed by multiplex networks (MNs) which consist of a fixed set of nodes connected by various sets of edges called layers \cite{Boccaletti14,Gao22,Lee14,Lee15}. In the simplest case, the layers are independently generated random networks with a full overlap of nodes, i.e., with each node belonging to all layers, which means it has at least one attached edge from each layer. In turn, in MNs with partial overlap of nodes, there are nodes belonging only to some rather than all layers. In particular, in the case of MNs with two layers (duplex networks) and partial overlap of nodes, the nodes are divided into a class of nodes belonging to both layers and forming the overlap, and two other classes, each consisting of nodes belonging only to one of the two layers \cite{Buono14, Alvarez19, Perez20} (the node overlap should not be confused with the link overlap \cite{Bianconi13, Min15, Diakonova16} which is negligible in the case of independently generated layers). FM phase transition in equilibrium models on MNs was studied, e.g., in the Ising model \cite{Krawiecki17, Krawiecki19} and a related Ashkin-Teller model \cite{Jang15}. Analogously, FM transition in nonequilibrium models for the opinion formation on MNs was studied, e.g., in the majority vote model \cite{Krawiecki18, Choi19}, the $q$-voter model \cite{Chmiel15, Gradowski20, Chmiel20} and the $q$-neighbor Ising model \cite{Chmiel17}. As expected, the critical properties of the nonequilibrium models, in particular the extension or confinement of the range of parameters for which the first-order transition occurs, strongly depend on the way in which the multiplexity affects the spin-flip rate. In this respect, very interesting seems the $q$-neighbor Ising model with LOCAL\&AND spin update rule \cite{Lee14a}, which so far has been studied by MC simulations and in the MF approximation on duplex networks with full and partial overlap of nodes and with layers in the form of fully connected graphs \cite{Chmiel17}. In this model, the flip probability per unit time for the spins in nodes belonging to only one layer (i.e., outside the overlap) is given by a Metropolis-like rate, but with a local field produced only by a subset of $q$ randomly chosen neighboring spins ($q$-neighborhood), and for the spins in nodes belonging to both layers (i.e., within the overlap) it is given by a product of two above-mentioned rates evaluated separately for each layer. With the increase of the relative size of the overlap, and depending on the size of the $q$-neighborhood, suppression of the first-order transition, appearance of a tricritical point separating first- and second-order FM transition, and possible coexistence of the PM and FM phases even in zero temperature were observed in the model \cite{Chmiel17}. 

In this paper, the $q$-neighbor Ising model on MNs with partial overlap of nodes, with layers in the form of complex networks and with the LOCAL\&AND spin update rule is studied by means of MC simulations and theoretically in the framework of the PA and AMEs. It should be noted that the $q$-neighbor Ising model is used here as a convenient example since the results can be readily compared with the above-mentioned ones for the limiting case of the model on MNs with layers in the form of complete graphs \cite{Chmiel17}, and the PA and AMEs used here can be easily generalized to other models for the opinion formation with similar structure of interactions. In order to make large systems of AMEs numerically tractable in this paper only the case of duplex networks with layers in the form of homogeneous random networks is considered; nevertheless, such MNs exhibit certain overlap-induced inhomogeneity since the nodes within and outside the overlap form distinct classes characterized by different degrees within the individual layers (both non-zero or one zero and one non-zero). Thus also the flip rates for the spins located in nodes belonging to distinct classes are different; a related $q$-voter model with quenched disorder, with agents divided into subpopulations according to different rates of the opinion change, has been recently considered in Ref.\ \cite{Jedrzejewski22}.

The aim of this paper is first to provide a general formulation of the PA and AMEs, which take into account to a different extent the above-mentioned inhomogeneity of nodes, for models on MNs with partial overlap of nodes. For this purpose, first, the homogeneous PA for models on MNs with a full overlap of nodes \cite{Gradowski20} is extended to the case with partial overlap. For nodes belonging to different classes this simplest form of the PA takes into account the inhomogeneity of the average directions of spins (opinions) but neglects possible inhomogeneity of the distributions of directions of neighboring spins within each layer. For the $q$-neighbor Ising model predictions of this approximation concerning the FM phase transition show surprisingly good agreement with results of MC simulations for a wide range of the size of the $q$-neighborhood, the mean degrees of nodes within layers and the size of the overlap. Then, the most advanced approximation based on the AMEs for models on MNs with the full overlap of nodes \cite{Choi19} and weighted networks \cite{Unicomb18} is extended to the case of models on MNs with partial overlap of nodes. Finally, two kinds of heterogeneous PA, the fully heterogeneous PA \cite{Jedrzejewski22} and the AMEs-based heterogeneous PA \cite{Gleeson11, Gleeson13, Peralta20} are applied to models on MNs with partial overlap of nodes. Both versions of the PA take into account, to a different extent, the above-mentioned inhomogeneity of distributions of directions of neighboring spins within each layer and are in general intermediate with respect to the accuracy of predictions between the homogeneous PA and the AMEs. For the $q$-neighbor Ising model under study, it turns out that their predictions are only marginally different or even identical with these of the homogeneous PA. On the other hand, predictions based on the AMEs show slightly better quantitative agreement with the results of MC simulations, in particular for smaller mean degrees of nodes within layers. In general, predictions of all approximations concerning the first-order FM transition (e.g., location and width of the hysteresis loop) are quantitatively worse than those concerning the second-order transition (e.g., location of the critical point). Besides, the aim of this paper is also to study in detail the phase diagram for the $q$-neighbor Ising model on MNs with partial overlap of nodes and with layers with a finite mean degree of nodes. It is shown that the critical behavior of this model resembles qualitatively that of the analogous model on MNs with layers in the form of fully connected graphs \cite{Chmiel17}. However, as the mean degree of nodes is decreased, the first-order FM transition, the tricritical point separating it from the second-order transition, and the possible coexistence of the PM and FM phases occur for smaller relative sizes of the overlap, while the range of the occurrence of the second-order FM transition is broadened correspondingly. 

%           ----
%            ||
%            ||
%            ||
%           ----

\section{The model}
\label{model}

\subsection{Multiplex networks with partial overlap of nodes}
\label{multnet}

MNs consist of a fixed set of nodes connected by several sets of edges; the set of nodes with each set of edges forms a network which is called a layer of a MN \cite{Boccaletti14, Gao22, Lee14, Lee15}. Henceforth, the nodes are indexed by $i$, $i=1,2,\ldots N$, and the subsequent layers are denoted as $G^{(L)}$, $L=A,B,\ldots L_{\max}$. In the case of MNs with a full overlap of nodes each node belongs to all layers, i.e., each node has at least one edge from each layer attached to it. In general, MNs with partial overlap of nodes are defined as MNs in which nodes may belong to (i.e., may have attached edges from) some rather than all layers, given that each node belongs to at least one layer. Henceforth, the number of nodes belonging to the layer $G^{(L)}$ is denoted as $N^{(L)}$. In this paper, it is assumed that the sets of edges for the subsequent layers $G^{(L)}$ are generated independently and form complex random networks with $N^{(L)}$ nodes. As a result, multiple connections between nodes are not allowed within the same layer, but the same nodes belonging to several layers can be accidentally connected by multiple edges belonging to different layers. A simple example of the MN with partial overlap of nodes is that with only two layers $G^{(A)}$, $G^{(B)}$, called a duplex network, and with $n$ nodes belonging to both layers which form the overlap ($0\le n\le N$); then, $N=N^{(A)}+N^{(B)}-n$. Furthermore, if both layers contain the same number of nodes $N^{(A)}=N^{(B)}=\tilde{N}$ it is possible to introduce a single parameter $r=n/\tilde{N}$, also called the overlap. Then, the nodes are divided into three subsets: $\tilde{N}-n= N (1-r)/(2-r)$ nodes belonging only to the layer $G^{(A)}$, $N (1-r)/(2-r)$ nodes belonging only to the layer $G^{(B)}$ and $n=Nr/(2-r)$ nodes belonging both to $G^{(A)}$ and $G^{(B)}$. 

The numbers of edges attached to the node $j$ (degrees) within the individual layers $G^{(L)}$ are denoted as $k_{j}^{(L)}$; if the node $j$ does not belong to the layer $G^{(L)}$ then $k_{j}^{(L)}=0$. In the case of MNs with independently generated layers the degrees of nodes belonging to the individual layers $G^{(L)}$, i.e., these with $k_{j}^{(L)}>0$, are drawn from probability distributions $P^{(L)}\left( k^{(L)}\right)$ which characterize the layers as complex networks. For a given node $j$ a vector of its degrees within the individual layers $\mathbf{k}_{j}=\left( k_{j}^{(A)}, k_{j}^{(B)},\ldots k_{j}^{(L_{\max})} \right)$, with possible zero components in the case of MNs with partial overlap of nodes, is called a multidegree of the node. The multidegree distribution $P(\mathbf{k}) = P\left( k^{(A)}, k^{(B)},\ldots k^{(L_{\max})} \right)$ characterizes the MN as a complex "network of networks"; in the case of MNs with the full overlap of nodes and independently generated layers, it is obviously $P(\mathbf{k})= \prod_{L=A}^{L_{\max}} P^{(L)}(k^{(L)})$. In the formulas below, averages are evaluated over the multidegree distribution, e.g., $\langle k^{(L)}\rangle$ $=N^{-1}\sum_{j=1}^{N} k_{j}^{(L)}$ $=\sum_{\mathbf{k}} P(\mathbf{k})k^{(L)}$ is the mean degree of nodes within the layer $G^{(L)}$ (note that the average is over all $N$ nodes rather than $N^{(L)}$ nodes belonging to the layer $G^{(L)}$). 

In this paper, the $q$-neighbor Ising model is considered on a duplex network with partial overlap of nodes and with two independently generated layers characterized by identical degree distributions $P^{(A)}(k)=P^{(B)}(k)=\tilde{P}(k)$, with the same numbers of nodes $N^{(A)}=N^{(B)}=\tilde{N}$ and the overlap $r$, for which the multidegree distribution is
\begin{equation}
\label{JointPDgen}
P\left( \mathbf{k}\right) = P\left( k^{(A)}, k^{(B)}\right) =
\frac{1-r}{2-r} \tilde{P}(k^{(A)})\delta_{k^{(B)},0}+ \frac{r}{2-r} \tilde{P}(k^{(A)})\tilde{P}(k^{(B)}) +
\frac{1-r}{2-r} \delta_{k^{(A)},0}\tilde{P}(k^{(B)}). 
\end{equation}
In particular, for the two layers in the form of random regular graphs (RRGs) with $K$ edges attached to each node belonging to the layer, for which $P^{(A)}(k)=P^{(B)}(k)=\tilde{P}(k)=\delta_{k,K}$, the multidegree distribution is
\begin{equation}
\label{JointPD}
P\left( \mathbf{k}\right) = P\left( k^{(A)}, k^{(B)}\right) =
\frac{1-r}{2-r} \delta_{k^{(A)},K}\delta_{k^{(B)},0}+ \frac{r}{2-r} \delta_{k^{(A)},K}\delta_{k^{(B)},K} +
\frac{1-r}{2-r} \delta_{k^{(A)},0}\delta_{k^{(B)},K},
\end{equation}
and $\langle k^{(A)}\rangle =\langle k^{(B)}\rangle =\tilde{N}K/N= K/(2-r)$. 

\subsection{The $q$-neighbor Ising model on multiplex networks with partial overlap of nodes}
\label{model_det}

The $q$-neighbor Ising model \cite{Jedrzejewski15, Park17, Jedrzejewski17a, Chmiel18, Chmiel17} is a nonequilibrium variant of the Ising model used to investigate the process of opinion formation. In this paper the above-mentioned model is considered on MNs with partial overlap of nodes and layers in the form of complex networks; the MF version of this model, on MNs with layers in the form of fully connected graphs, was studied in Ref.\ \cite{Chmiel17}. The main interest is in the FM transition which can occur in the $q$-neighbor Ising model with decreasing effective temperature $T$, which measures the level of internal noise (uncertainty in agents' decision making).

In order to introduce the model under study, it is convenient to start with the $q$-neighbor Ising model on (monoplex) networks which can be regular, complex, or fully connected graphs \cite{Jedrzejewski15, Park17, Jedrzejewski17a, Chmiel18, Chmiel17}. In this model agents with two possible opinions on a given subject are represented by two-state spins $\sigma_{j}=\pm 1$, $j=1,2,\ldots N$ placed in the nodes and interacting via edges of the network. It is assumed that these interactions prefer identical orientations of spins in the connected nodes, which is reflected in the spin-flip rate. Thus, interactions between spins with opposite directions in general increase the probability that one of the spins flips, i.e., the corresponding agent changes opinion, and edges representing these interactions are called active links. The dynamics of the $q$-neighbor Ising model on networks is a modification of that of the kinetic Ising model with the Metropolis spin-flip rate in which, at each time step, each spin interacts only with its $q$ randomly chosen neighbors. MC simulations of the model are performed using random asynchronous updating of spins, with each MC simulation step (MCSS) corresponding to updating all $N$ spins. Nodes are picked randomly and for each picked node $q$ its neighbors are chosen randomly and without repetitions, which form the $q$-neighborhood of the picked node. Then, the spin in the picked node is flipped with probability given by a Metropolis-like formula,
\begin{equation}
\label{M1}
    E\left( l;T,q\right) = \min \left\{ 1, \exp [-2 (q-2l)/T]\right\},
\end{equation}
where $l$ is the number of nodes belonging to the $q$-neighborhood occupied by spins with a direction opposite to that of the spin in the picked node, i.e., the number of active links attached to the picked node leading to nodes within the chosen $q$-neighborhood (notation in Eq.\ (\ref{M1}) emphasizes that $T$, $q$, are parameters of the model). As a result, the flip rate for a picked spin given that it is placed in a node with degree $k$ which has in total $i$ active links attached ($0\le i\le k$) is
\begin{equation}
f\left( i;T|k\right) = \frac{1}{{k \choose q}} \sum_{l=0}^{q} {i \choose l} {k-i \choose q-l} E\left( l;T,q\right)
=  \frac{1}{{k \choose i}} \sum_{l=0}^{q} {k-q \choose i-l} {q \choose l}E\left( l;T,q\right).
\label{fk}
\end{equation}

The $q$-neighbor Ising model on complete graphs for $q=3$ exhibits second-order FM transition, while for $q\ge 4$ first-order FM transition occurs with a clearly visible hysteresis loop. Width of the hysteresis loop in general increases with $q$, though for $q>4$ there are oscillations superimposed on this trend such that loops for the consecutive odd values of $q$ are narrower than for the neighboring even values of $q$ \cite{Jedrzejewski15}. The same is true for the model on networks with finite mean degree $\langle k \rangle$ provided that $q\ll \langle k\rangle$. However, as $q$ is increased and becomes comparable with $\langle k\rangle$ the hysteresis loop becomes narrower and eventually disappears, and the FM transition becomes second-order \cite{Chmiel18}. 

In the $q$-neighbor Ising model on MNs with full or partial overlap of nodes, interactions take place within individual layers with respective, independently chosen $q$-neighborhoods. Then, spins flip according to a probabilistic rule which combines the effect of the above-mentioned interactions. In this paper, the LOCAL\&AND spin update rule is used \cite{Lee14a} according to which the spin in the picked node flips if interaction with every $q$-neighborhood from every layer it belongs to suggests flip; consequently, the probability of the spin-flip is given by a product of the Metropolis-like factors (\ref{M1}) corresponding to all layers to which the picked node belongs. The LOCAL\&AND rule is assumed in this paper since it usually leads to richer phase diagrams than other methods of including the multiplex character of the network of interactions in the spin-flip rate \cite{Chmiel15, Gradowski20, Chmiel20, Chmiel17}. Eventually, in numerical simulations of the $q$-neighbor Ising model on MNs with partial overlap of nodes and the LOCAL\&AND spin update rule, each MCSS is performed as follows.
\begin{itemize}
    \item[(i.)] A node $j$, $1\le j \le N$, with multidegree $\mathbf{k}_i$ is picked randomly.
    \item[(ii.)] From each layer $G^{(L)}$ to which the picked node belongs a set of its $q$ neighbors ($q$-neighborhood) is chosen randomly and without repetitions. It is assumed that $0<q\le k_{j}^{(L)}$ for all layers to which the picked node belongs; otherwise, the node is excluded from simulation. Sets from different layers are chosen independently, thus the same node can by chance belong to two or more $q$-neighborhoods if it is a neighbor of the picked node within two or more layers.
    \item[(iii.)] The Metropolis-like factor for the picked node is evaluated separately for each layer $G^{(L)}$,
    \begin{equation}
E\left( l^{(L)};T,q\right) = \min \left\{ 1, \exp [-2 (q-2l^{(L)})/T]\right\}
\label{M2}
\end{equation}
where $l^{(L)}$ is the number of nodes in the $q$-neighborhood in the layer $G^{(L)}$ occupied by spins with direction opposite to that of the spin in the picked node; note that if a node does not belong to $G^{(L)}$ then $q=l^{(L)}=0$ and $E(T,0,0)=1$.
\item[(iv.)] Due to the LOCAL\&AND spin update rule, the spin $\sigma_j$ in the picked node flips with probability
\begin{equation}
    E\left( \mathbf{l};T,q\right) = \prod_{L=A}^{L_{\max}} E\left( l^{(L)}; T,q\right),
\end{equation}
where $\mathbf{l}=\left( l^{(A)},l^{(B)},\ldots l^{(L_{\max})}\right);$ and obviously $l^{(L)}=0$ if the picked node does not belong to the layer $G^{(L)}$ (i.e., $\mathbf{l}$ is a vector of numbers of active links from the individual layers attached to the picked node which lead to nodes within the respective $q$-neighborhoods).
\item[(v.)] Steps (i.)-(iv.) are repeated until all $N$ spins are updated without repetition.
\end{itemize}
Hence, the flip rate for a spin placed in a node with multidegree $\mathbf{k}=\left( k^{(A)}, k^{(B)},\ldots k^{(L_{\max})} \right)$ and with the numbers of attached active links within the individual layers $i^{(L)}$, $0\le i^{(L)}\le k^{(L)}$, given by the corresponding components of the vector $\mathbf{i}=\left( i^{(A)}, i^{(B)},\ldots i^{(L_{\max})} \right)$ assumes a multiplicative form,
\begin{equation}
    f\left( \mathbf{i};T\left| \mathbf{k}\right.\right)= \prod_{L=A}^{L_{\max}} f\left( i^{(L)};T|k^{(L)}\right)
    %=\prod_{L=A}^{L=L_{\max}} \frac{1}{{k^{(L)} \choose i^{(L)}}} \sum_{l^{(L)}=0}^{q} {k^{(L)}-q \choose i^{(L)}-l^{(L)}} {q \choose l^{(L)}}E\left( l^{(L)}; T,q\right)
\label{fkMN}
\end{equation}
(note that if a node does not belong to the layer $G^{(L)}$ there is $k^{(L)}=i^{(L)}=0$ and $f\left( 0;T|0\right)\equiv 1$).

The $q$-neighbor Ising model on a duplex network with layers in the form of complete graphs and partial overlap of nodes, and with the LOCAL\&AND spin update rule was investigated in Ref\ \cite{Chmiel17}. The obtained results may be summarized as follows. The model exhibits FM phase transition already for $q\ge 1$ \cite{Chmiel17}. This transition is in general second-order, with the following exceptions. For $q=2$ the transition is first-order for $1/2< r< 1$, with a clearly visible hysteresis loop, and for $r_c <r \le 1/2$, where $r_c=2(3\sqrt{2}-4)=0.4853\ldots$, the coexistence of the FM and PM phases is observed as the temperature is decreased below a critical value down to $T=0$; for $r<r_c$ there is no phase transition and the PM phase remains the only stable phase down to $T=0$. For $q\ge 4$ the transition for small $r$ is first-order and for larger $r$ is second-order. The first- and second-order transitions are separated by a tricritical point at $r=r_{TCP}(q)$ which for $q=4$ occurs at a particularly high value of $r$, and for $q>4$ is an increasing function of $q$, but again with oscillations between the consecutive odd and even values of $q$ superimposed on this trend. Remarkably, for $r=1$ the FM transition is always second-order for any $q$, i.e., a full overlap of nodes suppresses discontinuous transition. In this paper, it is investigated how the phase diagram of the model changes if the layers of the MN are complex networks with a finite mean degree of nodes rather than complete graphs.

%           -----
%            |||
%            |||
%            |||
%           -----

\section{Theory}
\label{theory}

\subsection{Pair approximation}
\label{HoPA}

In the case of spin models on networks, the effect of the network topology (e.g, of the degree distribution or the mean degree of nodes) on the observed phase transitions often can be more accurately described in the framework of the PA than by the usual MFA \cite{Vazquez08, Pugliese09, Gleeson11, Gleeson13, Peralta20}. In particular, this was demonstrated for the $q$-neighbor Ising model on complex networks \cite{Chmiel18} and a sort of stochastic $q$-voter model on MNs with a full overlap of nodes \cite{Gradowski20}. In both above-mentioned studies the networks, or the layers of the MNs, were homogeneous complex networks (e.g., RRGs), thus the simplest homogeneous PA was enough to reproduce quantitatively results of MC simulations in a wide range of the parameters of the models. As mentioned in Sec.\ \ref{intro} \& \ref{model} MNs with partial overlap of nodes retain some multiplexity-induced inhomogeneity even if the layers are homogeneous complex networks. Nevertheless, in this section the homogeneous PA derived in Ref.\ \cite{Gradowski20} for a wide class of models with various spin update rules on MNs with the full overlap of nodes is presented in a more general form which makes it applicable to models on MNs with partial overlap of nodes, in order to find, inter alia, to what extent it can be used to explain critical behavior of systems with multiplexity-induced inhomogeneity. 

The advantage of the PA consists in that it takes into account dynamical correlations between pairs of interacting agents (spins). In the framework of the homogeneous PA, macroscopic quantities characterizing a model with two-state spins on MNs are concentrations $c_{\mathbf{k}}$ of spins directed up located in nodes with multidegree $\mathbf{k}$ (with possible zero components in the case of MNs with partial overlap of nodes) as well as concentrations $b^{(L)}$ of active links within separate layers $G^{(L)}$. The homogeneous character of the PA allows for the simplification that the latter concentrations are averaged over all nodes belonging to a given layer and do not depend on the multidegrees of the connected nodes. Consequently, it is assumed that conditional probabilities $\theta_{\nu}^{(L)}$, $\nu\in \left\{ \uparrow,\downarrow\right\}$, that an active link within the layer $G^{(L)}$ is attached to a node given that it is occupied by spin with direction $\nu$ are also independent of the multidegree of the node. These probabilities can be evaluated as ratios of the number of attachments of active links to nodes with spins with direction $\nu$, independently of their multidegrees, within the layer $G^{(L)}$, which is $N\langle k^{(L)} \rangle b^{(L)}/2$, and the number of attachments of all links within $G^{L}$ to such nodes, which is $\sum_{\mathbf{k}} NP\left( \mathbf{k}\right)k^{(L)} c_{\mathbf{k},\nu}$, where  $c_{\mathbf{k},\uparrow}= c_{\mathbf{k}}$, $c_{\mathbf{k},\downarrow}= 1-c_{\mathbf{k}}$, thus
\begin{eqnarray}
    \theta_{\uparrow}^{(L)}&=&\frac{b^{(L)}}{2\sum_{\mathbf{k}} P\left( \mathbf{k}\right)k^{(L)} c_{\mathbf{k},\uparrow}/\langle k^{(L)} \rangle} =\frac{b^{(L)}}{2\sum_{\mathbf{k}} P\left( \mathbf{k}\right)k^{(L)} c_{\mathbf{k}}/\langle k^{(L)} \rangle},\label{condprobspinsu} \\
    \theta_{\downarrow}^{(L)}&=&\frac{b^{(L)}}{2\sum_{\mathbf{k}} P\left( \mathbf{k}\right)k^{(L)} c_{\mathbf{k},\downarrow}/\langle k^{(L)} \rangle}=\frac{b^{(L)}}{2\left[1- \sum_{\mathbf{k}} P\left( \mathbf{k}\right)k^{(L)} c_{\mathbf{k}}/\langle k^{(L)} \rangle\right]}
\label{condprobspinsd}
\end{eqnarray}  

The core approximation made in the PA for models on MNs is that the numbers of active links $i^{(L)}$ attached to a node with degrees $k^{(L)}$ within individual layers $G^{(L)}$ ($0\le i^{(L)}\le k^{(L)}$) occupied by spin with direction $\nu$ obey independent binomial distributions with parameters $\theta_{\nu}^{(L)}$ given by Eq.\ (\ref{condprobspinsu},\ref{condprobspinsd}). Then, the rates at which the concentration $c_{\mathbf{k}}$ increases or decreases are given by averages of the spin-flip rate, Eq.\ (\ref{fkMN}), over the appropriate joint distributions of the number of active links within all layers which have a multiplicative form 
\begin{equation}
P(\nu,\mathbf{i}|\mathbf{k}) =\prod_{L=A}^{L_{\max}} B_{k^{(L)},i^{(L)}}\left( \theta_{\nu}^{(L)}\right),
\label{bindist}
\end{equation}
where $B_{k,i}( \theta) = {k \choose i} \theta^{i} (1-\theta)^{k-i}$ denotes the binomial factor and, formally, $B_{0,0}(\theta) \equiv 1$. Hence, the equation for the time dependence of $c_{\mathbf{k}}$ can be written as a rate equation,
\begin{widetext}
\begin{eqnarray}
\frac{\partial c_{\mathbf{k}}}{\partial t} &=& \sum_{\nu\in \left\{ \uparrow,\downarrow\right\}} (-1)^{\delta_{\nu,\uparrow}}c_{\mathbf{k},\nu}
\sum_{\mathbf{i}} \prod_{L=A}^{L_{\max}}
 B_{k^{(L)},i^{(L)}}\left( \theta_{\nu}^{(L)}\right) 
f\left( \mathbf{i};T\left| \mathbf{k}\right.\right),
\label{rate1}
\end{eqnarray}
where $\sum_{\mathbf{i}} \equiv \sum_{i^{(A)}=0}^{k^{(A)}} \ldots \sum_{i^{(L_{\max})}=0}^{k^{(L_{\max})}}$.

In order to obtain an equation for the time dependence of the concentrations of active links $b^{(L)}$ one should observe that each flip of a spin (irrespective of its direction) in a picked node with multidegree $\mathbf{k}$ with the numbers of active links attached given by the components of the vector $\mathbf{i}$ results in the change of the numbers of active links within the individual layers $G^{(L)}$ by $k^{(L)}-2i^{(L)}$, since then $i^{(L)}$ previously active links become inactive and $k^{(L)}-i^{(L)}$ previously inactive links become active. The corresponding changes in the concentrations of active links $b^{(L)}$ are thus $\left( k^{(L)}-2i^{(L)}\right)/(N\langle k^{(L)}\rangle/2)$. As in Eq.\ (\ref{rate1}), such changes connected with the flip of a spin with direction $\nu$ occur at a rate given by the average of the spin-flip rate, Eq.\ (\ref{fkMN}), over the appropriate joint distributions of the number of active links attached to the picked node, Eq.\ (\ref{bindist}). Due to the homogeneous character of the PA, in order to obtain time dependence of $b^{(L)}$ further averaging over all nodes occupied by spins with direction $\nu$ should be performed, which is equivalent to averaging over the probability distribution $P(\mathbf{k}) c_{\mathbf{k},\nu}$ that a node with multidegree $\mathbf{k}$ is occupied by a spin with direction $\nu$. Eventually, taking into account that nodes are picked and spins are updated within time intervals $1/N$, for a given layer $G^{(L')}$ it is obtained that
\begin{eqnarray}
\frac{\partial b^{(L')}}{\partial t} &=& \frac{2}{\langle k^{(L')}\rangle} \sum_{\nu\in \left\{ \uparrow,\downarrow\right\}} 
\sum_{\mathbf{k}} P\left( \mathbf{k}\right) 
c_{\mathbf{k},\nu}
\sum_{\mathbf{i}} \prod_{L=A}^{L_{\max}}
 B_{k^{(L)},i^{(L)}}\left( \theta_{\nu}^{(L)}\right) 
f\left( \mathbf{i};T\left| \mathbf{k}\right.\right) \left( k^{(L')}-2i^{(L')}\right), \label{rate2} 
\label{ratebL}
\end{eqnarray}
\end{widetext}
where $L'=A,B\ldots L_{\max}$.

In particular, let us consider the $q$-neighbor Ising model on a MN with two layers in the form of RRGs and partial overlap of nodes, with the multidegree distribution given by Eq.\ (\ref{JointPD}). Then, the nodes are divided into three classes, these belonging only to the layer $G^{(A)}$ with multidegree $\mathbf{k}=(K,0)$, only to the layer $G^{(B)}$ with $\mathbf{k}=(0,K)$ and to the overlapping part of $G^{(A)}$ and $G^{(B)}$, with $\mathbf{k}=(K,K)$. The macroscopic quantities to be used in the homogeneous PA are thus concentrations of spins directed up in the nodes belonging to the subsequent classes $c_{(K,0)}$, $c_{(0,K)}$, $c_{(K,K)}$ and concentrations of active links in the two layers $b^{(A)}$, $b^{(B)}$. Since both layers are identical, with $N^{(A)}=N^{(B)}=\tilde{N}$, stable solutions of the system of equations (\ref{rate1}), (\ref{ratebL}) are limited to the subspace with $c_{(0,K)}=c_{(K,0)}$, $b^{(A)}=b^{(B)}\equiv b$; moreover, according to Eq.\ (\ref{condprobspinsu},\ref{condprobspinsd}) there is $\theta^{(A)}_{\nu}=\theta^{(B)}_{\nu}\equiv \theta_{\nu}$. Using Eq.\ (\ref{JointPD}), (\ref{fk}), (\ref{fkMN}), performing summations in Eq.\ (\ref{rate1}), (\ref{ratebL}) as in Ref.\ \cite{Chmiel18} and introducing functions $R(\theta;T,q)$ and $S(\theta;T,K,q)$ to shorten notation,
\begin{eqnarray}
R(\theta;T,q) &=& \sum_{l=0}^{q} B_{q,l}\left( \theta\right) E(l;T,q) 
\label{fun_R}, \\
S(\theta;T,K,q) &=& \sum_{l=0}^{q} B_{q,l}\left( \theta\right) [(K-q)\theta +l] E(l;T,q),
\label{fun_S}
\end{eqnarray}
the following system of three equations for the time dependence of the macroscopic quantities in the homogeneous PA is obtained,
\begin{eqnarray}
\frac{d c_{(K,0)}}{dt} &=& 
\left( 1-c_{(K,0)}\right) R\left(\theta_{\downarrow};T,q\right)
-c_{(K,0)} R\left(\theta_{\uparrow};T,q\right) 
\label{dcK0dt_hom}\\
\frac{d c_{(K,K)}}{dt} &=& 
\left( 1-c_{(K,K)}\right) \left[ R\left(\theta_{\downarrow};T,q\right) \right]^2
-c_{(K,K)}\left[R\left(\theta_{\uparrow};T,q\right) \right]^2
\label{dcKKdt_hom}\\
\frac{d b}{dt} &=&
\frac{2}{K} (1-r) \left\{
\left( 1-c_{(K,0)}\right)\left[ KR\left(\theta_{\downarrow};T,q\right)
-2S\left(\theta_{\downarrow};T,K,q\right)\right]  + c_{(K,0)}\left[ KR\left(\theta_{\uparrow};T,q\right)
-2S\left(\theta_{\uparrow};T,K,q\right)\right] \right\} \nonumber\\
&+& \frac{2}{K}r \left\{
\left( 1-c_{(K,K)}\right)\left[ KR\left(\theta_{\downarrow};T,q\right)
-2S\left(\theta_{\downarrow};T,K,q\right)\right]
R\left(\theta_{\downarrow};T,q\right)\right. \nonumber\\
&& \;\;\;\;\;\;\; +\left. c_{(K,K)}\left[ KR\left(\theta_{\uparrow};T,q\right)
-2S\left(\theta_{\uparrow};T,K,q\right)\right]
R\left(\theta_{\uparrow};T,q\right) \right\},
\label{dbdt_hom}
\end{eqnarray}
where 
\begin{eqnarray}
    \theta_{\uparrow} &=& \frac{b}{2\left[(1-r)c_{(K,0)} +r c_{(K,K)}\right]}, \\
    \theta_{\downarrow} &=& \frac{b}{2\left[1-(1-r)c_{(K,0)} -r c_{(K,K)}\right]}.
\end{eqnarray}
Other macroscopic quantities of interest are the concentration of spins directed up in each layer, i.e., the fraction of $\tilde{N}$ nodes occupied by such spins, which is $\tilde{c}=(1-r)c_{(K,0)}+rc_{(K,K)}$, the concentration of spins directed up in the whole MN, i.e., the fraction of $N$ nodes occupied by such spins, which is $c=\frac{2(1-r)}{2-r}c_{(K,0)}+\frac{r}{2-r} c_{(K,K)}$, and the resulting magnetization of the MN $m=2c-1$. Note that in the limiting case of layers in the form of fully connected graphs there is $b=\tilde{N}^2 \tilde{c}(1-\tilde{c})/[\tilde{N}(\tilde{N}-1)/2]\approx 2\tilde{c}(1-\tilde{c})$ and $\theta_{\downarrow}=\tilde{c}$, $\theta_{\uparrow}=1-\tilde{c}$; after inserting this into Eq.\ (\ref{dcK0dt_hom}) and (\ref{dcKKdt_hom}) equations for the concentrations $c_{(K,0)}$, $c_{(K,K)}$ in the MF approximation are reproduced \cite{Chmiel17}, as expected.

The above result can be easily extended to a more general case of the $q$-neighbor Ising model on a MN with two independently generated layers with identical degree distributions and numbers of nodes, and with the overlap $r$, where the multidegree distribution given by Eq.\ (\ref{JointPDgen}). Since due to the particular form of the spin-flip rate (\ref{fk}) the function $R(\theta;T,q)$, Eq.\ (\ref{fun_R}), does not depend on the multidegree distribution, equations (\ref{rate1}) for the concentrations $c_{k,0}$ for any $k>0$ will have the form of Eq.\ (\ref{dcK0dt_hom}), and for the concentrations $c_{k,k'}$ for any $k,k'>0$ will have the form of Eq.\ (\ref{dcKKdt_hom}).  Hence, the stationary solution is characterized only by two, possibly different, concentrations of spins directed up
$c_{k,k'}$, $c_{k,0}$ in nodes belonging or not to the overlap, respectively. Then, as a result of averaging over $P(\bf{k})$, the equation (\ref{ratebL}) for the concentration $b$ of active links within each layer will have the form of Eq.\ (\ref{dbdt_hom}) with $K$ replaced by the average degree of nodes with respect to the probability distribution within individual layers,
$\tilde{\langle k \rangle} = \sum_{k}\tilde{P}(k)k$. Eventually, the homogeneous PA for the $q$-neighbor Ising model on MNs with partial overlap of nodes is not sensitive to the details of the multidegree distribution $P(\bf{k})$ apart from the mean degree within layers $\tilde{\langle k \rangle}$, as for the $q$-neighbor Ising model on networks \cite{Chmiel18}. Thus, predictions of the homogeneous PA do not change even if the layers of the MN are heterogeneous (e.g.\, scale-free) networks rather than homogeneous RRGs. In fact, it is known that MC simulations of the $q$-neighbor Ising model \cite{Chmiel18} and of the related noisy $q$-voter model \cite{Jedrzejewski17,Jedrzejewski22} on networks with various heterogeneity, including scale-free networks, yield quantitatively similar results. This suggests that also the critical properties of the $q$-neighbor Ising model on MNs with partial overlap of nodes should not depend on the heterogeneity of the degree distributions of nodes within individual layers, and the accuracy of predictions of the homogeneous PA should be comparable with that for the $q$-neighbor Ising model on MNs with partial overlap of nodes and layers in the form of RRGs. The study of this issue is left for future research.

A natural extension of the homogeneous PA consists in taking into account heterogeneity of the concentrations of the (possibly active) links connecting classes of nodes with different multidegrees, so that, instead of the average concentration $b^{(L)}$ of active links within the layer $G^{(L)}$, e.g., concentrations of classes of active links connecting spins in nodes with multidegrees ${\bf k}$, ${\bf k}'$ within the layer $G^{(L)}$ become separate macroscopic quantities characterizing the model. This leads to the most advanced and accurate version of the PA called fully heterogeneous PA \cite{Jedrzejewski22, Pugliese09}; corresponding equations for the macroscopic quantities for spin models on MNs with partial overlap of nodes, in particular for the $q$-neighbor Ising model under study, are given in Appendix A. In the latter case solutions of these equations show that in the stationary state concentrations of active links (strictly speaking, of their ends called bonds) belonging to different classes indeed show noticeable heterogeneity; nevertheless, this does not lead to the values of magnetization
noticeably different from these predicted by the homogeneous PA. Thus, magnetization curves and phase diagrams for the model under study obtained from the fully heterogeneous PA are practically indistinguishable from those obtained from the homogeneous PA and do not show better agreement with the results of MC simulations.

\subsection{Approximate Master equations}
\label{AMEs}

A more accurate approximation for the study of spin models on MNs with partial overlap of nodes is based on approximate Master equations (AMEs) for the densities of spins directed up $c_{\mathbf{k,m}}$ and down $s_{\mathbf{k,m}}$ which are located in nodes with multidegree $\mathbf{k}$ and have $m^{(L)}$ neighboring spins directed up within the consecutive layers $G^{(L)}$, which is denoted as $\mathbf{m}=\left( m^{(A)}, m^{(B)}\ldots m^{(L_{\max})}\right)$. In the thermodynamic limit and for mutually uncorrelated layers in the form of random networks with finite mean degrees $\big \langle k^{(L)}\big \rangle$ possibility that a pair of nodes is connected simultaneously by edges within different layers can be neglected. Thus, in the AMEs it is assumed that  in a single simulation step for a given node the allowed changes of the number of neighboring spins directed up are $\mathbf{m}\rightarrow \mathbf{m}\pm \mathbf{e}^{(L)}$, where $\mathbf{e}^{(L)}$ is a unit vector with $L_{\max}$ components and only $L$-th component equal to one, while simultaneous changes of many components of $\mathbf{m}$, e.g.,  $\mathbf{m}\rightarrow \mathbf{m}\pm \mathbf{e}^{(L)}\pm \mathbf{e}^{(L')}$, $L\neq L'$, etc., cannot occur. Under the above-mentioned assumptions, the AMEs in a general form are \cite{Unicomb18, Choi19}
\begin{eqnarray}
\frac{d s_{\mathbf{k,m}}}{dt} &=& - F_{\mathbf{k,m}} s_{\mathbf{k,m}}+ R_{\mathbf{k,m}} c_{\mathbf{k,m}} 
\nonumber\\
&+& \sum_{L=A}^{L_{\max}}\left[ - \beta_{s}^{(L)} \left(k^{(L)}-m^{(L)}\right)  s_{\mathbf{k,m}}
 + \beta_{s}^{(L)} \left(k^{(L)}-m^{(L)}+1\right)  s_{\mathbf{k,m}-\mathbf{e}^{(L)}} \right]
\nonumber\\
&+& \sum_{L=A}^{L_{\max}}\left[ - \gamma_{s}^{(L)} m^{(L)} s_{\mathbf{k,m}}
 + \gamma_{s}^{(L)} \left(m^{(L)}+1\right)  s_{\mathbf{k,m}+\mathbf{e}^{(L)}} \right],
\label{dskmdt}\\
\frac{d c_{\mathbf{k,m}}}{dt} &=& - R_{\mathbf{k,m}} c_{\mathbf{k,m}}+ F_{\mathbf{k,m}} s_{\mathbf{k,m}} 
\nonumber\\ 
&+& \sum_{L=A}^{L_{\max}}\left[ - \beta_{c}^{(L)} \left(k^{(L)}-m^{(L)}\right)  c_{\mathbf{k,m}}
 + \beta_{c}^{(L)} \left(k^{(L)}-m^{(L)}+1\right)  c_{\mathbf{k,m}-\mathbf{e}^{(L)}} \right]
\nonumber\\
&+& \sum_{L=A}^{L_{\max}}\left[ - \gamma_{c}^{(L)} m^{(L)} c_{\mathbf{k,m}}
 + \gamma_{c}^{(L)} \left(m^{(L)}+1\right)  c_{\mathbf{k,m}+\mathbf{e}^{(L)}} \right].
\label{dikmdt}
\end{eqnarray}
In Eq.\ (\ref{dskmdt}), (\ref{dikmdt}) the first two terms account for the effect of a flip of a spin in a node with 
multidegree $\mathbf{k}$ and the remaining terms account for the average effect of the flips of spins in the neighboring nodes, irrespective of their multidegrees. In terms of Sec.\ \ref{model_det} the flip rate for a spin directed down occupying a node with multidegree $\mathbf{k}$ with $\mathbf{m}$ neighboring spins directed up is $F_{\mathbf{k,m}}= f \left( \mathbf{m}; T \left| \mathbf{k} \right. \right)$ and that for a spin directed
up $R_{\mathbf{k,m}}= f \left( \mathbf{k-m}; T \left| \mathbf{k} \right. \right)$. The remaining average rates can be estimated by evaluating the ratios (at a given time step)
of the average number of edges connecting
spins with a given direction such that one of these spins flips to the average total numbers of these edges \cite{Gleeson11, Gleeson13}; in the case of models on MNs this should be done separately for each layer
\cite{Unicomb18,Choi19}. Thus
$\beta_{s}^{(L)} =\big \langle\sum_{\mathbf{m}}\left( k^{(L)}-m^{(L)} \right) F_{\mathbf{k,m}} s_{\mathbf{k,m}}\big \rangle/
\big \langle\sum_{\mathbf{m}}\left( k^{(L)}-m^{(L)} \right) s_{\mathbf{k,m}}\big \rangle$, 
$\gamma_{s}^{(L)} = \big \langle\sum_{\mathbf{m}}\left( k^{(L)}-m^{(L)} \right) R_{\mathbf{k,m}} c_{\mathbf{k,m}}\big \rangle/$ 
$\big \langle\sum_{\mathbf{m}}\left( k^{(L)}-m^{(L)} \right) c_{\mathbf{k,m}}\big \rangle$, \hspace{0.3cm}
$\beta_{c}^{(L)} = \big \langle\sum_{\mathbf{m}} m^{(L)} F_{\mathbf{k,m}} s_{\mathbf{k,m}}\big \rangle/
\big \langle\sum_{\mathbf{m}}m^{(L)} s_{\mathbf{k,m}}\big \rangle$, \hspace{0.3cm}
$\gamma_{c}^{(L)} = \big \langle\sum_{\mathbf{m}} m^{(L)} R_{\mathbf{k,m}} c_{\mathbf{k,m}}
\big \rangle/$ 
 $\big \langle\sum_{\mathbf{m}} m^{(L)} c_{\mathbf{k,m}}\big \rangle$,  where $L=A,B,\ldots L_{\max}$,
$\sum_{\mathbf{m}} \equiv \sum_{m^{(A)}=0}^{k^{(A)}} \sum_{m^{(B)}=0}^{k^{(B)}} \ldots \sum_{m^{(L_{\max})}=0}^{k^{(L_{\max})}} $ and
$\langle\ldots \rangle$ denotes average over the multidegree distribution $P\left( \mathbf{k}\right)$, as usually. Natural initial conditions for the system of equations (\ref{dskmdt}), (\ref{dikmdt}) are $s_{\mathbf{k,m}}(0) = (1-c(0))\prod_{L=A}^{L_{\max}} B_{k^{(L)},m^{(L)}}(c(0))$, $c_{\mathbf{k,m}}(0) = c(0)\prod_{L=A}^{L_{\max}} B_{k^{(L)},m^{(L)}}(c(0))$, where $0< c(0) <1$ is arbitrary. 

In particular, in the case of the $q$-neighbor Ising model on a MN with two layers in the form of RRGs and partial overlap of nodes, with the multidegree distribution $P(\mathbf{k})$ given by Eq.\ (\ref{JointPD}), there are three classes of nodes with $\mathbf{k}=(0,K)$, $\mathbf{k}=(K,0)$ and $\mathbf{k}=(K,K)$. The corresponding spin flip rates are
$F_{(K,0),\left(m^{(A)};0\right)}= f \left( m^{(A)};T \left| K \right. \right)$,
$F_{(0,K),\left(0;m^{(B)}\right)}= f \left( m^{(B)}; T \left| K \right. \right)$,
$F_{(K,K),\left(m^{(A)},m^{(B)}\right)}= f \left( m^{(A)}; T \left| K \right. \right) f \left( m^{(B)}; T \left| K \right. \right)$ and
$R_{(K,0),\left(m^{(A)},0\right)}= f \left(K- m^{(A)};T \left| K \right. \right)$,
$R_{(0,K),\left(0,m^{(B)}\right)}= f \left( K-m^{(B)}; T \left| K \right. \right)$,
$R_{(K,K),\left(m^{(A)},m^{(B)}\right)}= f \left( K-m^{(A)}; T \left| K \right. \right) f \left(K-m^{(B)}; T \left| K \right. \right)$, with $f \left( m; T \left| K \right. \right)$ given by Eq.\ (\ref{fk}). Hence, the system (\ref{dskmdt}), (\ref{dikmdt}) consists of $2(K+1)^2+4(K+1)$ equations and can be solved numerically for moderate $K$. The quantities of interest, e.g., the concentration $c$ of spins directed up in the MN and the magnetization $m=2c-1$ can be evaluated as in Sec.\ \ref{HoPA} using $c_{(K,0)}=\sum_{m^{(A)}=0}^{K} c_{(K,0),(m^{(A)},0)}$, $c_{(0,K)}=\sum_{m^{(B)}=0}^{K} c_{(0,0),(0,m^{(B)})}$, $c_{(K,K)}=\sum_{m^{(A)}=0}^{K} \sum_{m^{(B)}=0}^{K} c_{(K,K),(m^{(A)},m^{(B)})}$.

The AMEs are a starting point for a more elaborate approximation representing another formulation of the heterogeneous PA \cite{Gleeson11,Gleeson13,Peralta20,Choi19,Unicomb18} which takes into account the possible heterogeneity due to
different multidegrees $\mathbf{k}$ of nodes of both the concentrations $c_{\mathbf{k}}$ of spins directed up and
of the conditional probabilities that a link attached to a node is active or, equivalently, leads to a spin with a given (say, up) direction. A general formulation of such AMEs-based heterogeneous PA for spin (two-state) models on (monoplex) networks by Gleeson \cite{Gleeson11,Gleeson13} was extended to the case of weighted networks \cite{Unicomb18} and, partly, MNs \cite{Choi19}. It is believed that due to the approximations made the AMEs-based heterogeneous PA is in general more accurate than the homogeneous PA and less accurate than the fully heterogeneous PA mentioned in Sec. \ref{HoPA}. In this paper the AMEs-based heterogeneous PA is applied to spin models on MNs with partial overlap of nodes, in particular to the $q$-neighbor Ising model under study; equations for the macroscopic quantities are given in Appendix B. Surprisingly, in numerical simulations it turns out that in the stationary state the above-mentioned conditional probabilities that a node has a link leading to a spin directed up do not depend on whether the node belongs or not to the overlap. Hence, predictions of the AMEs-based heterogeneous PA concerning the FM transition in the model under study are identical to those of the homogeneous PA from Sec.\ \ref{HoPA}, so they are not further discussed.

\begin{figure}
    \includegraphics[width=\linewidth]{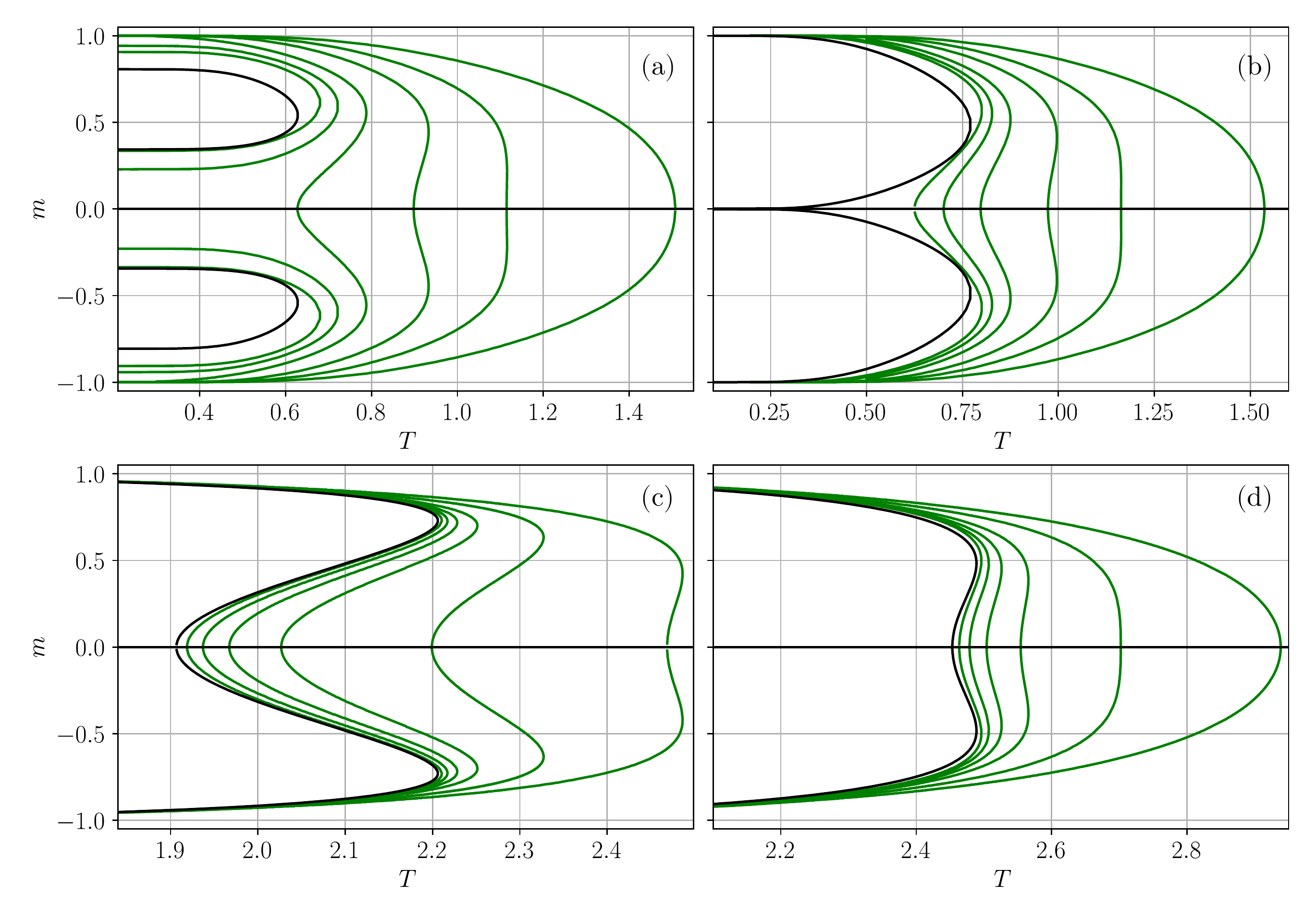}
    \caption{Magnetization $m$ vs.\ temperature $T$ predicted by the homogeneous PA for different $K$ (green solid lines, both stable and unstable fixed points of the system of equations (\ref{dcK0dt_hom}-\ref{dbdt_hom}) are shown) and from the MFA of Ref.\ \cite{Chmiel17} (black solid lines) for $q=2$, $K=200,100,50,20,10,4$ (from left) and (a) $r=0.49$, (b) $r=0.5$; as well as for $q=4$, $K=500,200,100,50,20,10$ (from left) and (c) $r=0.05$, (d) $r=0.15$.}
    \label{fig:mf_pa}
\end{figure}

\begin{figure}
    \includegraphics[width=\linewidth]{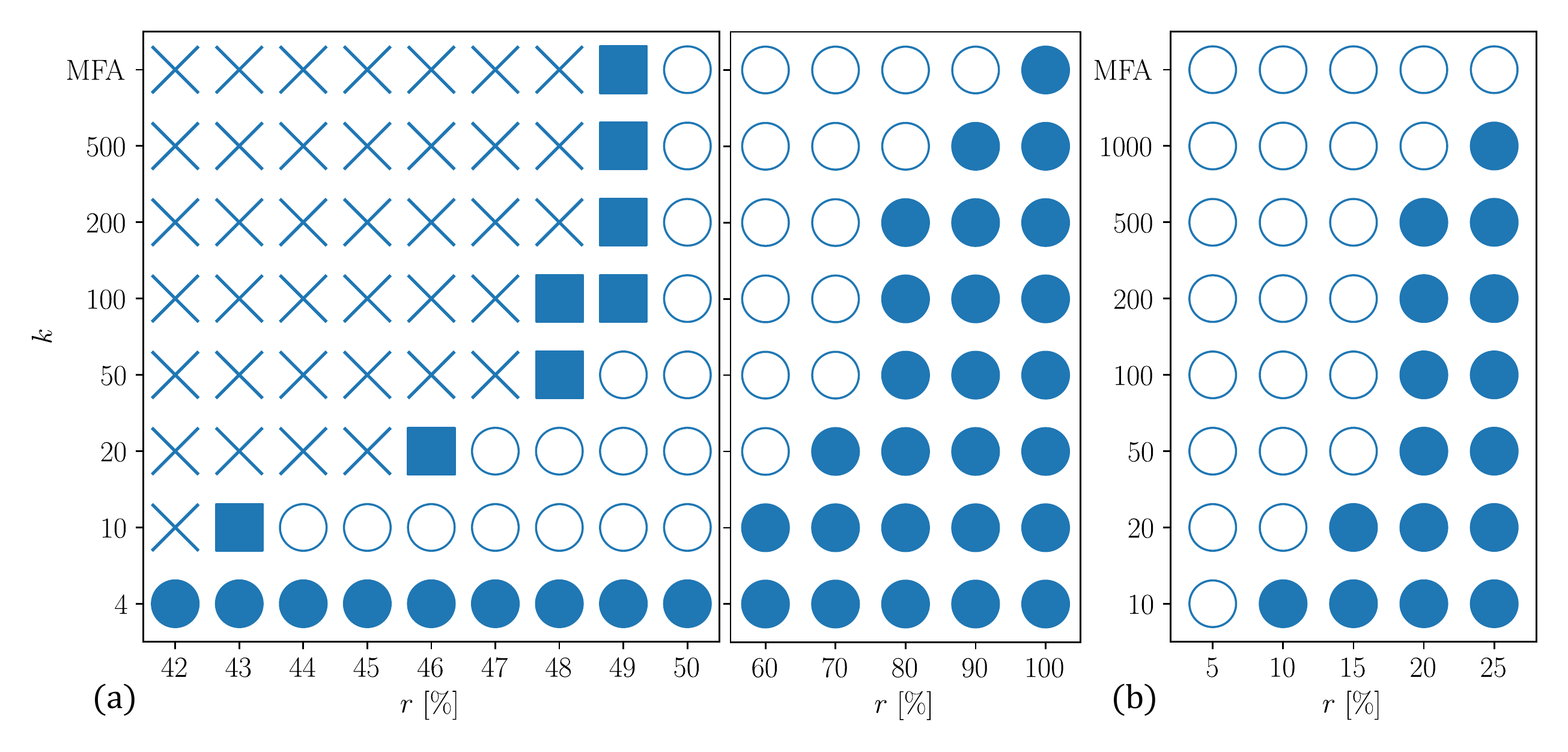}
    \caption{Critical behavior predicted by the homogeneous PA for the model with (a) $q=2$ (left and middle panels) and (b) $q=4$ (right panel) and different $r$, $K$; filled circles --- continuous FM transition, open circles --- discontinuous FM transition, filled squares --- coexistence of the FM and PM phases for $T\rightarrow 0$, crosses --- absence of the transition.}
    \label{fig:pa_k_r}
\end{figure}

\begin{figure}
    \includegraphics[width=\linewidth]{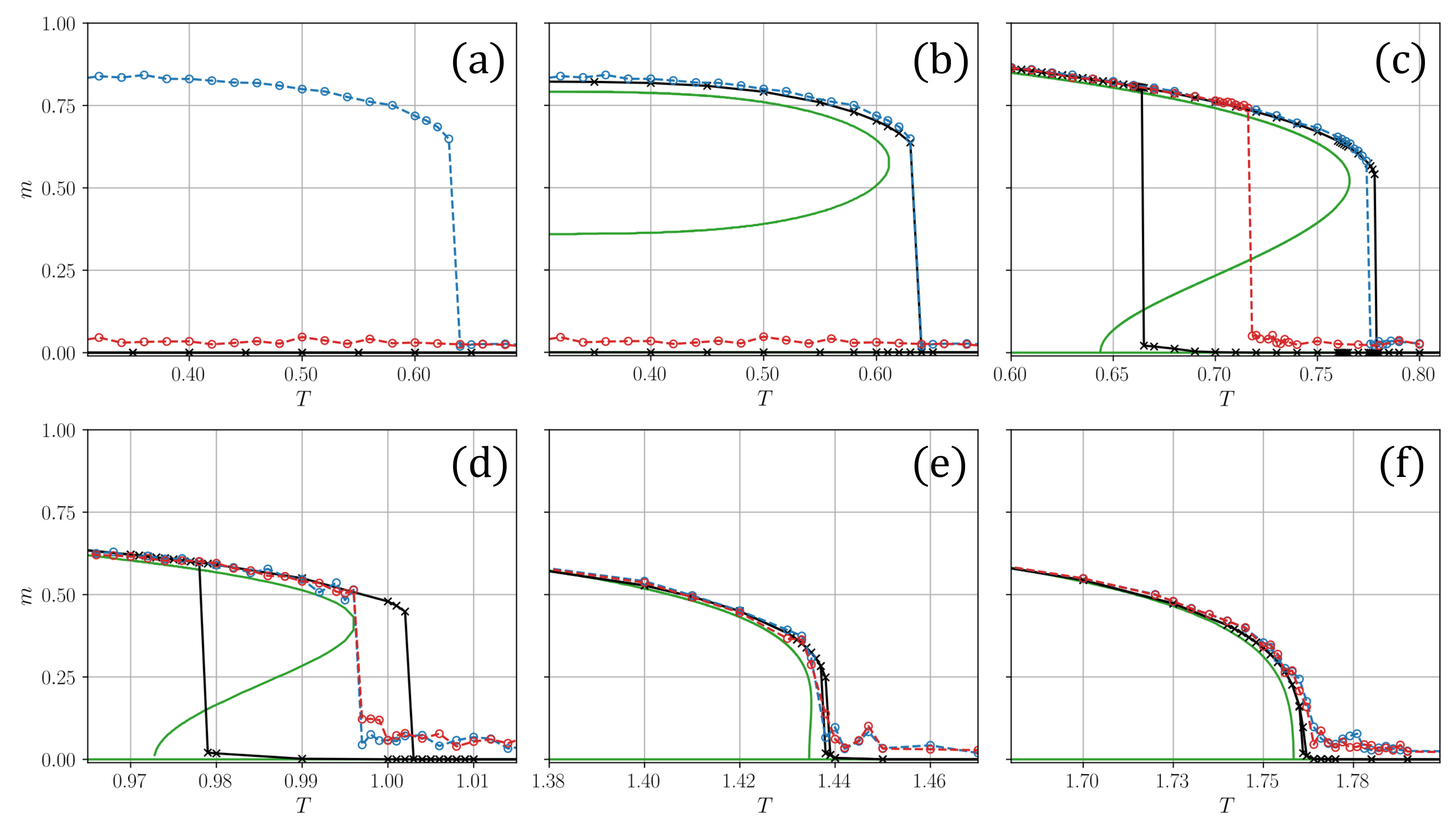}
    \caption{Results of MC simulations as well as predictions of the PA and AMEs for the model with $q=2$, $K=20$ and (a) $r=0.45$, (b) $r=0.46$, (c) $r=0.47$, (d) $r=0.50$, (e) $r=0.60$, (f) $r=0.70$; blue dots --- results of MC simulations with FM initial conditions and increasing temperature, red dots --- results of MC simulations with PM initial conditions and decreasing temperature, black dots --- predictions of the AMEs for both FM ($c(0)=1$) and PM ($c(0)=0.5$) initial conditions and increasing or decreasing temperature, respectively, green solid lines --- predictions of the PA as in Fig.\ \ref{fig:mf_pa}. }
    \label{fig:mT_q2} 
\end{figure}

\begin{figure}
    \includegraphics[width=\linewidth]{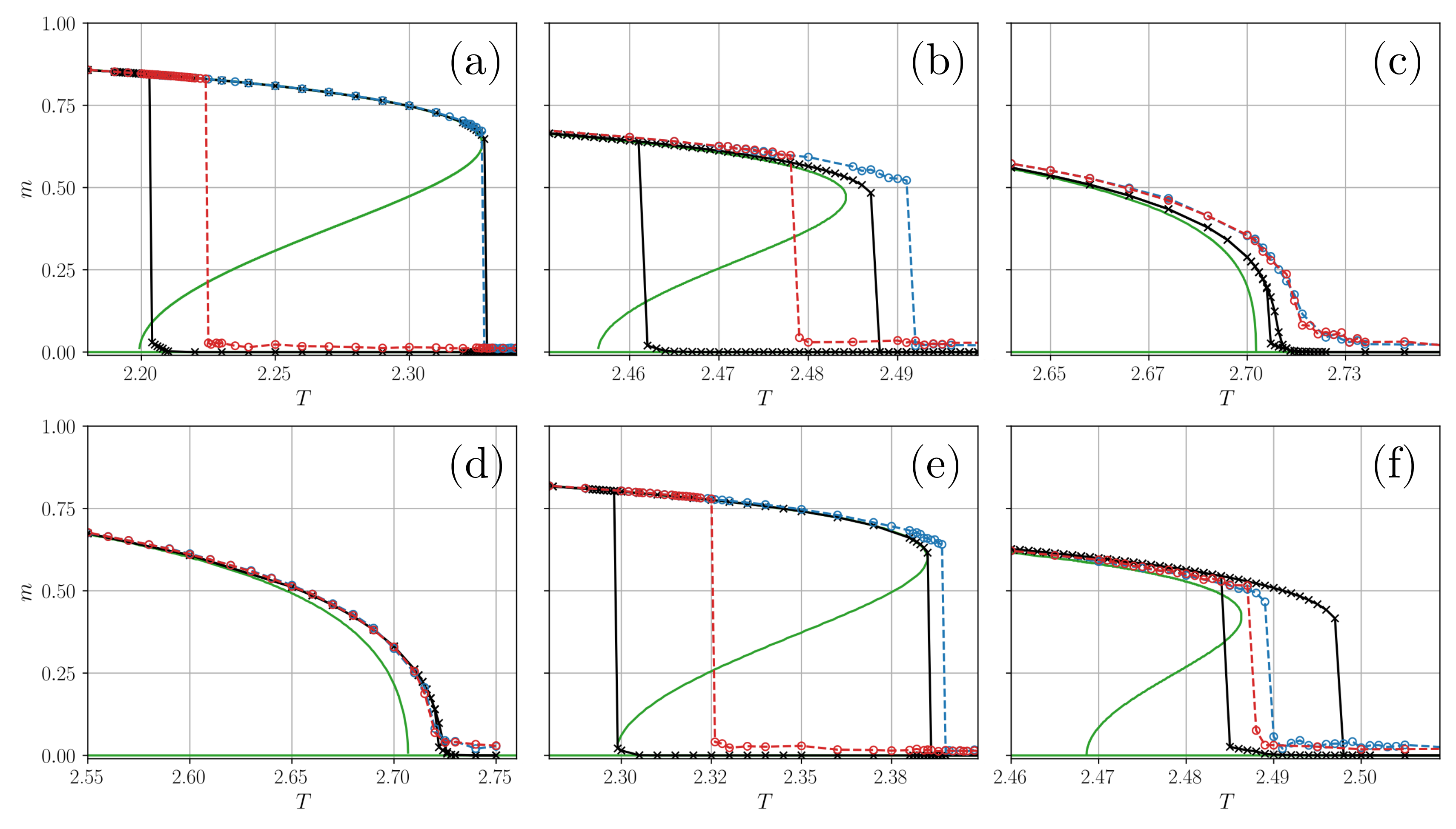}
    \caption{As in Fig.\ \ref{fig:mT_q2} but for $q=4$. (a) $K=20$, $r=0.05$, (b) $K=20$, $r=0.10$, (c) $K=20$, $r=0.15$, (d) $K=10$, $r=0.10$, (e) $K=50$, $r=0.10$, (f) $K=10$, $r=0.05$.}
    \label{fig:mT_q4}
\end{figure}

%           --------
%            | |   |
%            |  | |
%            |   |
%           --------

\section{Results}
\label{Res}

The main results concerning the FM transition in the $q$-neighbor Ising model on MNs with partial overlap of nodes and with layers in the form of complete graphs have been summarized in Sec.\ \ref{model_det}. These results were obtained in the MF approximation and confirmed by MC simulations \cite{Chmiel17}. In this section first predictions of the homogeneous PA of Sec.\ \ref{HoPA} concerning the FM transition in the $q$-neighbor Ising model on MNs with partial overlap of nodes and with layers in the form of RRGs are presented and compared with results of MC simulations. In this case, noticeable discrepancies occur between theoretical and numerical results, in particular concerning the first-order FM transition. As pointed out in Sec.\ \ref{theory}, the more advanced fully and AMEs-based heterogeneous PA yield results practically indistinguishable or even identical to the homogeneous PA, thus their predictions are only briefly mentioned in the Appendix. Finally, it is verified in which cases and to what extent theoretical predictions are improved by using the AMEs of Sec.\ \ref{AMEs}.  

In the framework of the homogeneous PA of Sec.\ \ref{HoPA} stationary values of the magnetization $m$ vs.\ $T$, corresponding to different thermodynamic phases, are given by stable fixed points of the system of equations (\ref{dcK0dt_hom}-\ref{dbdt_hom}) with $\dot{c}_{(K,0)}= \dot{c}_{(K,K)}=\dot{b}=0$. For certain ranges of parameters $r,q,K$ many stable fixed points can coexist for given $T$, and their basins of attraction are then separated by stable manifolds of unstable fixed points; location of the stable and unstable fixed points for the three-dimensional system of equations (\ref{dcK0dt_hom}-\ref{dbdt_hom}) can be determined using standard numerical tools. The homogeneous PA predicts various critical behavior of the model under study as the temperature $T$ is varied, depending on $r,q,K$ which are fixed: first- and second-order FM phase transition, the coexistence of the PM and FM phases for $T\rightarrow 0$ and absence of the FM transition. At high temperatures, the only stable fixed point is that with $m=0$ corresponding to the PM phase. In the case of the second-order FM transition, this fixed point loses stability as the temperature is decreased below the critical value $T_c$, and simultaneously a pair of symmetric stable fixed points with $m>0$ and $m<0$ occurs via a supercritical pitchfork bifurcation, corresponding to the two symmetric FM phases. In the case of the first-order transition two symmetric pairs of stable and unstable fixed points with $m>0$ and $m<0$ occur simultaneously via two saddle-node bifurcations as the temperature is decreased below the upper critical value $T_{c}^{(2)}$, and the two above-mentioned stable fixed points correspond to the two symmetric FM phases. As the temperature is further decreased both the FM and PM fixed points remain stable (coexist) until the PM point loses stability via a subcritical pitchfork bifurcation at the lower critical temperature $T_{c}^{(1)}$ ($T_{c}^{(1)}<T_{c}^{(2)}$) by colliding simultaneously with the two above-mentioned unstable fixed points; coexistence of the PM and FM phases for $T_{c}^{(1)}< T< T_{c}^{(2)}$ leads to the occurrence of the hysteresis loop in the magnetization curves $m(T)$. Eventually, for $T<T_{c}^{(1)}$ the only stable fixed points remain these corresponding to the two symmetric FM phases. In the case of the coexistence of the FM and PM phases for $T\rightarrow 0$ a pair of symmetric stable FM fixed points occurs at $T=T_{c}^{(2)}$ as in the case of the first-order transition, but these FM points, as well as the PM fixed point, remain stable (coexist) as $T\rightarrow 0$. Finally, it can also happen that fixed points corresponding to the FM phase do not exist for any $T>0$, thus the FM transition is absent and the only stable phase for $T\rightarrow 0$ is the PM one. 

Exemplary curves $m(T)$ predicted by the homogeneous PA for the model under study with different $K$ and selected values of $r$ are shown in Fig. \ref{fig:mf_pa} for the most interesting cases $q=2$ and $q=4$; in the former case, the MFA (valid for the model on MNs with layers in the form of fully connected graphs with $K\rightarrow \infty$) predicts occurrence of all above-mentioned kinds of the critical behavior for different ranges of $r$, while in the latter one it predicts occurrence of the first-order transition for a particularly wide range of small $r$. The curves $m(T)$ for $q=2$ are drawn in Fig. \ref{fig:mf_pa}(a) for $r=0.49$, and in Fig. \ref{fig:mf_pa}(b) for $r=0.5$, i.e., for the values of $r$ within or at the border of the interval $r_c<r<0.5$ where the MFA predicts coexistence of the FM and PM phases for $T\rightarrow 0$. In contrast, for the model on MNs with layers in the form of RRGs the homogeneous PA for $r=0.49$ (Fig. \ref{fig:mf_pa}(a)) predicts second- or first-order FM transition for small and moderate $K$, respectively; the critical temperature(s) decrease and the width of the hysteresis loop increases with $K$. Only for large $K$ coexistence of the FM and PM phases for $T\rightarrow 0$ is predicted by the PA, and the curves $m(T)$ approach those resulting from the MFA, as expected. For $r=0.5$ (Fig. \ref{fig:mf_pa}(b)) only second- or first-order FM transitions for finite $K$ are predicted by the PA, with the lower critical temperature  for the first-order transition $T_{c}^{(1)}>0$. The curves $m(T)$ for $q=4$ are drawn in Fig. \ref{fig:mf_pa}(c) for $r=0.05$, and in Fig. \ref{fig:mf_pa}(d) for $r=0.15$, i.e., for the values of $r$ where the MFA predicts first-order FM transition with a wide and narrow hysteresis loop, respectively. For the model on MNs with layers in the form of RRGs the homogeneous PA for small $r=0.05$ (Fig. \ref{fig:mf_pa}(c)) similarly predicts the first-order FM transition for moderate and large $K$, while for larger $r=0.15$ (Fig. \ref{fig:mf_pa}(d)) it predicts the second-order FM transition already for moderate $K$ and the first-order FM transition only for large $K$. Again, the critical temperature(s) decrease, and the width of the hysteresis loop increases with $K$, and the curves $m(T)$ eventually approach these resulting from the MFA. 

It may be inferred from Fig. \ref{fig:mf_pa} that the homogeneous PA predicts for the $q$-neighbor Ising model on MNs with partial overlap of nodes and layers in the form of RRGs with finite $K$ the same critical behavior as the MFA for the model on analogous MNs with layers in the form of complete graphs, only for different ranges of the overlap $r$. This conclusion is supported by Fig. \ref{fig:pa_k_r} where the critical behavior predicted by the PA is summarized for the former model with fixed $q= 2$ and $q=4$ and different $K$, $r$. For all $K$ and $r=1$ (full overlap of nodes), both PA and MFA predict continuous FM transition with decreasing $T$, i.e., the first-order transition observed in the model on monoplex networks is suppressed. However, for both $q=2,4$ and finite $K$ the PA predicts the second-order FM transition also for a range of $r$ below $r=1$ which is broadened with decreasing $K$. As a consequence, for $q=2$ (Fig. \ref{fig:pa_k_r}(a)) the PA predicts that the range of the occurrence of the first-order FM transition is shifted toward smaller values of $r$. Similarly, for a narrow range of still smaller values of the overlap the PA predicts the coexistence of the FM and PM phases for $T\rightarrow 0$, but for small $K$ this kind of critical behavior is predicted at $r$ significantly below the interval $r_c< r<0.5$ obtained from the MFA. Finally, it is predicted that the range of small $r$ for which the FM transition is absent for decreasing $K$ is narrowed. Eventually, for very small $K=4$ comparable with $q$ only continuous FM transition is predicted for any $r$, and all other kinds of critical behavior are suppressed. For $q=4$ (Fig. \ref{fig:pa_k_r}(b)) the range of small $r$ for which the PA predicts the first-order FM transition is substantially diminished with decreasing $K$.

In order to verify predictions of the homogeneous PA, MC simulations of the $q$-neighbor Ising model with $q=2,4$ on large MNs with various parameters $r$, $K$ were performed and the magnetization curves $m(T)$ were obtained for random PM initial conditions $\sigma_j =\pm 1$, $j=1,2,\ldots N$  and decreasing temperature as well as for FM initial conditions $\sigma_j =+1$, $j=1,2\ldots N$ and increasing temperature. Simulations were performed on MNs with $N^{(A)} = N^{(B)} = 10^5$. After setting the initial conditions, the system evolved for $10^4$ to $10^5$ MC steps to reach the stationary state. The magnetization was calculated as an average value over the next $10^3$ MC steps. Comparison with MC simulations shows that the homogeneous PA qualitatively captures modification of the critical behavior of the model under study due to finite values of the mean degree of the layers $K$. Quantitative predictions of the PA are much improved in comparison with those from the MFA valid for large $K$ and show reasonably good agreement with results of MC simulations, but
are not exact (Fig.\ \ref{fig:mT_q2}, \ref{fig:mT_q4}, note narrow ranges of $T$ covered by the horizontal axes). For fixed $q$, $K$ the PA approximately predicts the ranges of the overlap $r$ where different kinds of critical behavior should occur. However, as a rule, these predictions are overestimated and in MC simulations the particular kinds of critical behavior appear for smaller values of $r$ than estimated from the PA. For example, for $q=2$ the ranges of appearance of the coexistence of the FM and PM phases for $T\rightarrow 0$ (Fig. \ref{fig:mT_q2}(a,b)) and of the second-order FM transition (Fig. \ref{fig:mT_q2}(e,f)) in MC simulations are, respectively, shifted and extended toward smaller values of $r$ than expected from the PA. Consequently, in the case of the first-order FM transition for $q=2$ (Fig. \ref{fig:mT_q2}(c,d)) and $q=4$ (Fig. \ref{fig:mT_q4}(a,b,e,f)) the lower and upper critical temperatures $T_{c}^{(1)}$, $T_{c}^{(2)}$ are underestimated and the width of the hysteresis loop is overestimated by the PA in comparison with these obtained from MC simulations; it is interesting to note that discrepancies between the theoretical and numerical values of $T_{c}^{(2)}$ are usually smaller than those for $T_{c}^{(1)}$. Similarly, in the case of the second-order FM transition for $q=2$ (Fig. \ref{fig:mT_q2}(f)) and $q=4$ (Fig. \ref{fig:mT_q4}(c,d)) the critical temperature $T_c$ is underestimated by the PA in comparison with that obtained from MC simulations. In general, the curves $m(T)$ evaluated from the PA show better agreement with those obtained from MC simulations in the case of the second-order than the first-order FM transition. It should be mentioned that the discrepancies between the predictions of the homogeneous PA and results of the MC simulations are not a particular feature of the model under study; similar and even more significant discrepancies were observed in some cases for the $q$-neighbor Ising model on monoplex networks \cite{Chmiel18} as well as for the related noisy $q$-voter model on monoplex networks and MNs \cite{Jedrzejewski22,Gradowski20}.

In order to investigate the critical behavior of the model under study by means of appropriate AMEs as defined in Sec.\ \ref{AMEs}, Eq.\ (\ref{dskmdt},\ref{dikmdt}) were solved numerically with various initial conditions and the curves $m(T)$ were obtained using long-time asymptotic values of the concentrations of spins directed up $c_{(K,0),(m^{(A)},0)}$, etc., to evaluate stationary values of the magnetization. As expected, predictions of the AMEs usually show comparable or better agreement with the results of MC simulations than those of the homogeneous PA. This is particularly visible in the case of the second-order FM transition in the model under study with small $K$ and $q=2$ (Fig. \ref{fig:mT_q2}(f)) and $q=4$ (Fig. \ref{fig:mT_q4}(d)), where the theoretical and numerical curves $m(T)$ coincide very well and the critical temperature $T_c$ is predicted correctly. However, the ranges of the overlap predicted by the AMEs for which different kinds of critical behavior occur are still shifted toward slightly higher values of $r$ than obtained from MC simulations (see Fig. \ref{fig:mT_q2}(a), where the AMEs predict the absence of the FM transition rather than coexistence of the FM and PM phases observed in MC simulations, and Fig. \ref{fig:mT_q2}(e), where the AMEs predict the first-order FM transition with a narrow hysteresis loop rather than the second-order transition). In the case of the coexistence of the FM and PM phases for $T\rightarrow 0$ for $q=2$ (Fig. \ref{fig:mT_q2}(b)) and the first-order FM transition for $q=2$ (Fig. \ref{fig:mT_q2}(c,d)) and $q=4$ (Fig. \ref{fig:mT_q4}(a,b,e,f)) predictions of the AMEs concerning the upper critical temperature $T_{c}^{(2)}$ are usually better than those of the homogeneous PA, but the lower critical temperature $T_{c}^{(1)}$ is again usually underestimated and the width of the hysteresis loop is overestimated. In general, some improvement of theoretical predictions by the AMEs in comparison with the homogeneous PA can be seen for small and moderate $K$; for large $K$ the curves $m(T)$ obtained from the AMEs and PA coincide (Fig. \ref{fig:mT_q4}(e)). 

%           -----
%           |   |
%            | |
%             |
%           -----

\section{Discussion and conclusions}
\label{Concl}

In this paper the $q$-neighbor Ising model on MNs with partial overlap of nodes and layers in the form of random networks was investigated; as an example, the model on MNs with two layers in the form of RRGs was studied in detail. Both theoretical considerations based on the homogeneous PA and AMEs as well as MC simulations show that for given $q \ge 1$ and finite mean degree of nodes $K$, and for varying overlap $r$ and temperature $T$ the model exhibits qualitatively similar critical behavior as the $q$-neighbor Ising model on MNs with partial overlap of nodes and  layers in the form of complete graphs. In particular, for any $q$ and full overlap of nodes $r=1$ the first-order FM transition is suppressed and only the second-order transition appears with decreasing $T$. Besides, for decreasing $K$ continuous rather than discontinuous FM transition is observed for an increasing range of large (for $q=2$) and large and moderate (for $q >2$) values of $r$ below $r=1$. As a consequence, for decreasing $K$ the ranges of $r$ for which the model exhibits the first-order FM transition (for $q\ge 2$) and the coexistence of the FM and PM phases for $T\rightarrow 0$ (for $q=2$) are shifted toward smaller values. It should be mentioned that in the $q$-neighbor Ising model on (monoplex) networks the first-order FM transition is also suppressed for small $K$ comparable with $q$ \cite{Chmiel18}; in contrast, in the model on MNs this suppression is due to the overlap of nodes and occurs for any $K$. The $q$-neighbor Ising model was used here as an example, and related models for the opinion formation on MNs with partial overlap of nodes can be studied using similar numerical and analytic methods; however, the expected qualitative changes of the observed critical behavior with $r$ will be probably less spectacular, since, e.g., in the case of the $q$-voter model even for $r=1$ the first-order FM transition is not suppressed \cite{Gradowski20}. 

For the model under study with large $K$ predictions of the simple homogeneous PA and more advanced system of AMEs converge to these of the MFA and agree quantitatively with the results of MC simulations. For finite $K$ the predicted curves $m(T)$ and critical temperature(s) differ quantitatively from the numerically obtained ones; usually, the particular kinds of critical behavior are predicted to occur for smaller values of the overlap than observed in MC simulations. In general, predictions of both PA and AMEs show better agreement with the results of MC simulations in the case of the continuous than discontinuous FM transition. Predictions based on the AMEs are comparable to or better than those of the PA; in particular, the critical temperature for the second-order FM transition and the upper critical temperature for the first-order transition are more accurately predicted. Nevertheless, both PA and AMEs qualitatively correctly capture changes of the critical behavior of the model with varying parameters $K$, $r$ characterizing the underlying MN. 

Two versions of the heterogeneous PA were derived for the model under study, the more accurate fully heterogeneous PA and the less accurate AMEs-based heterogeneous PA, which take to a different extent into account heterogeneity of the distribution of the active links due to inhomogeneity of the nodes. In both cases, systems of equations for the macroscopic quantities characterizing the model are significantly larger and more complicated than that in the homogeneous PA but do not lead to a noticeable improvement of theoretical predictions concerning the magnetization curves and phase diagrams for the model. This suggests that the simple homogeneous PA can be as reliable as more advanced versions of the PA in the study of the critical behavior of systems with multiplexity-induced inhomogeneity, and conditions under which this is possible deserve further investigation. Only using much larger systems of AMEs in certain cases quantitatively improves agreement between theoretical predictions and results of MC simulations of the above-mentioned models.

% AAAAAAAAAAAAAAAAAAAAAAAAAAAAAAAAAAAAAAAAAAAAAAAAAAAAAAA%

\section*{Appendix}
\label{appendix}

\subsection{Fully heterogeneous pair approximation}

\label{appendixA}

The following outline of the fully heterogeneous PA for models on MNs is an extension of that for the $q$-voter models with quenched disorder on networks, with two populations of agents differing by the spin-flip rates \cite{Jedrzejewski22}. 
The fully heterogeneous PA uses the assumption that the probability that a spin directed up or down in a node with multidegree $\mathbf{k}$ has a given number of attached edges leading to spins directed up or down in nodes with multidegree $\tilde{\bf k}$ obeys a binomial distribution; this assumption is valid for each layer and for any pair ${\bf k}$, $\tilde{\bf k}$ separately, and the related binomial distributions are assumed to be independent. The macroscopic quantities characterizing a model with two-state spins on a MN are concentrations $c_{\bf k}$ of spins directed up in nodes with multidegree ${\bf k}$ and concentrations 
$e^{{\bf k},\tilde{{\bf k}}, (L)}_{\nu, \tilde{\nu}}$ 
($=e^{\tilde{\bf k},{\bf k}, (L)}_{\tilde{\nu}, \nu}$)
of bonds (ends of edges) attached within the layer $ G^{(L)}$ to nodes with multidegree ${\bf k}$ containing spins with direction $\nu\in \left\{ \downarrow, \uparrow\right\}$ such that at the other end of the edge there is a node with multidegree $\tilde{\bf k}$ containing spin with direction $\tilde{\nu}$ (normalized to the total number of bonds $N\langle k^{(L)}\rangle$ within $ G^{(L)}$). According to the above-mentioned assumptions the joint probability that a node with multidegree ${\bf k}$ containing spin with direction $\nu$ has ${\bf i}= \left( i^{(A)}, i^{(B)}, \ldots i^{(L_{\max})}\right)$
active bonds (ends of active links) attached within the consecutive layers, pointing at nodes with arbitrary multidegree containing spins with opposite direction $-\nu$, has a multiplicative form  
\begin{equation}
P(\nu,\mathbf{i}|\mathbf{k}) =\prod_{L=A}^{L_{\max}} B_{k^{(L)},i^{(L)}}\left( \alpha^{\mathbf{k},(L)}_{\nu}\right),
\end{equation}
where 
$\alpha^{\mathbf{k},(L)}_{\nu}=
\sum_{{\bf k}'} e^{{\bf k},{\bf k}', (L)}_{\nu, -\nu}/
\sum_{{\bf k}'}\sum_{\nu'\in \left\{ \downarrow, \uparrow\right\}} e^{{\bf k},{\bf k}', (L)}_{\nu, \nu'}$
are conditional probabilities that an active bond is attached to a node with multidegree ${\bf k}$ and spin with direction $\nu$ (similar to $\theta_{\uparrow}^{(L)}$, $\theta_{\downarrow}^{(L)}$ given by Eq.\ (\ref{condprobspinsu}, \ref{condprobspinsd})). In order to evaluate the change in the concentration 
$e^{{\bf k},\tilde{{\bf k}}, (L)}_{\nu, \tilde{\nu}}$ due to, e.g., flipping the spin with direction $\nu$ in a node with multidegree ${\bf k}$, it is necessary to know the numbers of bonds $y$, $z$ attached to this node within the layer $G^{(L)}$ pointing at nodes with multidegree $\tilde{\bf k}$ given that these bonds are active ($\tilde{\nu} =-\nu$) or inactive ($\tilde{\nu}=\nu$), respectively. These numbers obey binomial distributions $B_{i^{(L)},y}
\left( \beta^{{\bf k},\tilde{{\bf k}}, (L)}_{\nu, -\nu}\right)$,
$B_{k^{(L)}-i^{(L)},z}\left( \gamma^{{\bf k},\tilde{\bf k}, (L)}_{\nu, \nu}\right)$, respectively, where the conditional probabilities are 
$\beta^{{\bf k},\tilde{{\bf k}}, (L)}_{\nu, -\nu}=
e^{{\bf k},\tilde{{\bf k}}, (L)}_{\nu, -\nu}/
\sum_{{\bf k}'} e^{{\bf k},{\bf k}',(L)}_{\nu, -\nu}$,
$\gamma^{{\bf k},\tilde{{\bf k}}, (L)}_{\nu, \nu}=
e^{{\bf k},\tilde{{\bf k}}, (L)}_{\nu, \nu}/
\sum_{{\bf k}'} e^{{\bf k},{\bf k}',(L)}_{\nu, \nu}$. Then the rate equations for the macroscopic concentrations of spins directed up are 
\begin{eqnarray}
\frac{\partial c_{\mathbf{k}}}{\partial t} &=& \sum_{\nu\in \left\{ \uparrow,\downarrow\right\}} (-1)^{\delta_{\nu,\uparrow}}c_{\mathbf{k},\nu}
\sum_{\mathbf{i}} \prod_{L=A}^{L_{\max}}
 B_{k^{(L)},i^{(L)}}\left( \alpha^{\mathbf{k},(L)}_{\nu}\right) 
f\left( \mathbf{i};T\left| \mathbf{k}\right.\right),
\label{rate1hetero}
\end{eqnarray}
while the rate equations for the concentrations of active and inactive bonds within the layers contain terms such as
\begin{eqnarray}
\frac{d}{dt} e^{{\bf k},\tilde{{\bf k}}, (L')}_{\nu, -\nu} &=& 
\frac{1}{\langle k^{(L')}\rangle} 
P({\bf k}) c_{{\bf k},\nu} 
\sum_{\bf i} 
\prod_{L=A}^{L_{\max}}
B_{k^{(L)},i^{(L)}}\left( \alpha^{{\bf k},(L)}_{\nu}\right) 
\sum_{y=0}^{i^{(L')}}B_{i^{(L')},y}
\left( \beta^{{\bf k},\tilde{{\bf k}}, (L')}_{\nu, -\nu}\right) (-y)
f\left( \mathbf{i};T\left| \mathbf{k}\right.\right) + \ldots
 \label{rateeehetero}
\\
\frac{d}{dt} e^{{\bf k},\tilde{{\bf k}}, (L')}_{\nu, \nu} &=& 
\frac{1}{\langle k^{(L')}\rangle} 
P({\bf k}) c_{{\bf k},\nu} 
\sum_{\bf i} 
\prod_{L=A}^{L_{\max}}
B_{k^{(L)},i^{(L)}}\left( \alpha^{{\bf k},(L)}_{\nu}\right) 
\sum_{z=0}^{k^{(L')}-i^{(L')}}B_{k^{(L')}-i^{(L')},z}
\left( \gamma^{{\bf k},\tilde{{\bf k}}, (L')}_{\nu, \nu}\right) (-z)
f\left( \mathbf{i};T\left| \mathbf{k}\right.\right) +\ldots
\nonumber\\
\label{ratee-ehetero}
\end{eqnarray}
It can be seen that Eq.\ (\ref{rate1hetero}) resembles Eq.\ (\ref{rate1}) in the homogeneous PA. Concerning the equations for the concentrations of bonds, e.g., Eq.\ (\ref{rateeehetero}) states that a flip of the spin with direction $\nu$ in node with multidegree ${\bf k}$, which has $i^{(L')}$ active bonds attached within the layer $G^{(L')}$, out of which $y$ bonds point at nodes with multidegrees $\tilde{\bf k}$, decreases the concentration $e^{{\bf k},\tilde{{\bf k}}, (L')}_{\nu, -\nu}$ by $y/\left[ N\langle k^{(L')}\rangle\right]$; 
such a flip occurs with probability $P({\bf k}) c_{{\bf k},\nu} f\left( \mathbf{i};T\left| \mathbf{k}\right.\right)$ within a time interval $1/N$; and the final input to the rate equation (\ref{rateeehetero}) is obtained by averaging the above-mentioned change over the probability distributions $B_{k^{(L')},i^{(L')}}\left( \alpha^{{\bf k},(L)}_{\nu}\right)$
for $i^{(L')}$ and
$B_{i^{(L')},y} \left( \beta^{{\bf k},\tilde{{\bf k}}, (L')}_{\nu, -\nu}\right)$ for $y$; etc. Let us note that in Eq.\ (\ref{rateeehetero}), (\ref{ratee-ehetero}) averaging over the multidegree distribution $P(\bf{k})$ is not performed; this is in contrast with the homogeneous PA, Eq.\ ( \ref{ratebL}), as well as with the PA for models with two kinds of agents \cite{Jedrzejewski22}.

In the case of the $q$-neighbor Ising model on MNs with partial overlap of nodes and with two layers in the form of RRGs, with the multidegree distribution $P\left( \mathbf{k}\right)$ given by Eq.\ (\ref{JointPD}), there are three classes of nodes with $\mathbf{k}=(K,0)$, $\mathbf{k}=(0,K)$ and $\mathbf{k}=(K,K)$, and two layers $G^{(L)}$, $L=A,B$. Taking into account  the symmetry of the model under study and general symmetry conditions for the concentrations $e^{{\bf k},\tilde{{\bf k}}, (L)}_{\nu, \tilde{\nu}}$ the solutions of the system of equations (\ref{rate1hetero} - \ref{ratee-ehetero}) can be constrained to a 12-dimensional subspace
$c_{(K,0)}=c_{(0,K)}$, 
$e^{(K,0),(K,0),(A)}_{\;\;\;\;\nu\;\;\;,\;\;\;\nu'} =
e^{(K,0),(K,0),(A)}_{\;\;\;\;\nu'\;\;\;,\;\;\;\;\nu} =
e^{(0,K),(0,K),(B)}_{\;\;\;\;\nu\;\;\;,\;\;\;\;\nu'}=
e^{(0,K),(0,K),(B)}_{\;\;\;\;\nu'\;\;\;,\;\;\;\;\nu}\equiv
e^{(K,0),(K,0)}_{\;\;\;\;\nu\;\;\;,\;\;\;\;\nu'}$,
$e^{(K,0),(K,K),(A)}_{\;\;\;\;\nu\;\;\;,\;\;\;\;\nu'} =
e^{(K,K),(K,0),(A)}_{\;\;\;\;\nu'\;\;\;,\;\;\;\;\nu} =
e^{(0,K),(K,K),(B)}_{\;\;\;\;\nu\;\;\;,\;\;\;\;\nu'}=
e^{(K,K),(0,K),(B)}_{\;\;\;\;\nu'\;\;\;,\;\;\;\;\nu}\equiv
e^{(K,0),(K,K)}_{\;\;\;\;\nu\;\;\;,\;\;\;\;\nu'}$,
$e^{(K,K),(K,K),(A)}_{\;\;\;\;\nu\;\;\;,\;\;\;\;\nu'} =
e^{(K,K),(K,K),(A)}_{\;\;\;\;\nu'\;\;\;,\;\;\;\;\nu} =
e^{(K,K),(K,K),(B)}_{\;\;\;\;\nu\;\;\;,\;\;\;\;\nu'}=
e^{(K,K),(K,K),(B)}_{\;\;\;\;\nu'\;\;\;,\;\;\;\;\nu}\equiv
e^{(K,K),(K,K)}_{\;\;\;\;\nu\;\;\;,\;\;\;\;\nu'}$,
$\nu, \nu' \in \left\{ \downarrow, \uparrow \right\}$. 
Thus, as in Ref.\ \cite{Jedrzejewski22}, there are effectively only two classes of agents located in nodes with $\mathbf{k}=(K,0)$ and $\mathbf{k}=(K,K)$, differing by the spin-flip rates (\ref{fkMN}). Besides, the distributions of the number of links pointing at nodes belonging to each class given that these links are active or inactive are fully determined by the conditional probabilities   
$\beta^{{\bf k},{\bf k}}_{\nu, -\nu}$, $\gamma^{{\bf k},{\bf k}}_{\nu, \nu}$
for the links within each class. Taking this into account and
performing summations in Eq.\ (\ref{rate1hetero} - \ref{ratee-ehetero}) as in Ref.\ \cite{Chmiel18} the following system of equations for the macroscopic quantities is obtained in the fully heterogeneous PA for the model under study,
\begin{eqnarray}
\frac{d c_{(K,0)}}{dt} &=& 
\left( 1-c_{(K,0)}\right) R\left(\alpha^{(K,0)}_{\;\;\;\;\downarrow};T,q\right)
-c_{(K,0)} R\left(\alpha^{(K,0)}_{\;\;\;\;\uparrow};T,q\right) 
\label{dcK0dt_het}\\
\frac{d c_{(K,K)}}{dt} &=& 
\left( 1-c_{(K,K)}\right) \left[ R\left(\alpha^{(K,K)}_{\;\;\;\;\downarrow};T,q\right) \right]^2
-c_{(K,K)}\left[R\left(\alpha^{(K,K)}_{\;\;\;\;\uparrow};T,q\right) \right]^2
\label{dcKKdt_het}
\\
\frac{d}{dt} e^{(K,0),(K,0)}_{\;\;\;\;\uparrow\;\;\;,\;\;\;\;\uparrow} &=&
-\frac{2(1-r)}{K}c_{(K,0)} 
\gamma^{(K,0),(K,0)}_{\;\;\;\;\uparrow\;\;\;\;\,,\;\;\;\uparrow}
\left[KR \left( \alpha^{(K,0)}_{\;\;\;\;\uparrow};T,q\right)
-S\left( \alpha^{(K,0)}_{\;\;\;\;\uparrow};T,K,q\right) \right]
\nonumber\\
&+&
\frac{2(1-r)}{K}  \left( 1-c_{(K,0)}\right)
\beta^{(K,0),(K,0)}_{\;\;\;\;\downarrow\;\;\;\;\,,\;\;\;\uparrow}
S\left( \alpha^{(K,0)}_{\;\;\;\;\downarrow};T,K,q\right)
\\
\frac{d}{dt} e^{(K,0),(K,0)}_{\;\;\;\;\downarrow\;\;\;,\;\;\;\;\downarrow} &=&
\frac{2(1-r)}{K}c_{(K,0)} 
\beta^{(K,0),(K,0)}_{\;\;\;\;\uparrow\;\;\;\;\,,\;\;\;\downarrow}
S\left( \alpha^{(K,0)}_{\;\;\;\;\uparrow};T,K,q\right)
\nonumber\\
&-&
\frac{2(1-r)}{K}  \left( 1-c_{(K,0)}\right)
\gamma^{(K,0),(K,0)}_{\;\;\;\;\downarrow\;\;\;\;\,,\;\;\;\downarrow}
\left[KR \left( \alpha^{(K,0)}_{\;\;\;\;\downarrow};T,q\right)
-S\left( \alpha^{(K,0)}_{\;\;\;\;\downarrow};T,K,q\right) \right]
\\
\frac{d}{dt} e^{(K,K),(K,K)}_{\;\;\;\;\uparrow\;\;\;,\;\;\;\;\uparrow} &=&
-\frac{2r}{K}c_{(K,K)} 
\gamma^{(K,K),(K,K)}_{\;\;\;\;\uparrow\;\;\;\;\,,\;\;\;\;\uparrow}
\left[KR \left( \alpha^{(K,K)}_{\;\;\;\;\uparrow};T,q\right)
-S\left( \alpha^{(K,K)}_{\;\;\;\;\uparrow};T,K,q\right) \right]
R \left( \alpha^{(K,K)}_{\;\;\;\;\uparrow};T,q\right)
\nonumber\\
&+&
\frac{2r}{K}  \left( 1-c_{(K,K)}\right)
\beta^{(K,K),(K,K)}_{\;\;\;\;\downarrow\;\;\;\;\,,\;\;\;\;\uparrow}
S\left( \alpha^{(K,K)}_{\;\;\;\;\downarrow};T,K,q\right)
R\left( \alpha^{(K,K)}_{\;\;\;\;\downarrow};T,K,q\right)
\\
\frac{d}{dt} e^{(K,K),(K,K)}_{\;\;\;\;\downarrow\;\;\;,\;\;\;\;\downarrow} &=&
\frac{2r}{K}c_{(K,K)} 
\beta^{(K,K),(K,K)}_{\;\;\;\;\uparrow\;\;\;\;\,,\;\;\;\;\downarrow}
S\left( \alpha^{(K,K)}_{\;\;\;\;\uparrow};T,K,q\right)
R\left( \alpha^{(K,K)}_{\;\;\;\;\uparrow};T,K,q\right) \nonumber\\
&-&
\frac{2r}{K}  \left( 1-c_{(K,K)}\right)
\gamma^{(K,K),(K,K)}_{\;\;\;\;\downarrow\;\;\;\;\,,\;\;\;\;\downarrow}
\left[KR \left( \alpha^{(K,K)}_{\;\;\;\;\downarrow};T,q\right)
-S\left( \alpha^{(K,K)}_{\;\;\;\;\downarrow};T,K,q\right) \right]
R \left( \alpha^{(K,K)}_{\;\;\;\;\downarrow};T,q\right)
\\
\frac{d}{dt} e^{(K,0),(K,0)}_{\;\;\;\;\uparrow\;\;\;,\;\;\;\;\downarrow} &=&
\frac{1-r}{K}c_{(0,0)} 
\left\{
-\beta^{(K,0),(K,0)}_{\;\;\;\;\uparrow\;\;\;\;\,,\;\;\;\downarrow}
S\left( \alpha^{(K,0)}_{\;\;\;\;\uparrow};T,K,q\right)
\right. \nonumber\\
&+& 
\left. 
\gamma^{(K,0),(K,0)}_{\;\;\;\;\uparrow\;\;\;\;\,,\;\;\;\uparrow}
\left[KR \left( \alpha^{(K,0)}_{\;\;\;\;\uparrow};T,q\right)
-S\left( \alpha^{(K,0)}_{\;\;\;\;\uparrow};T,K,q\right) \right]
\right\}
\nonumber\\
&+&
\frac{1-r}{K}  \left( 1-c_{(K,0)}\right)
\left\{
\gamma^{(K,0),(K,0)}_{\;\;\;\;\downarrow\;\;\;\;\,,\;\;\;\downarrow}
\left[KR \left( \alpha^{(K,0)}_{\;\;\;\;\downarrow};T,q\right)
-S\left( \alpha^{(K,0)}_{\;\;\;\;\downarrow};T,K,q\right) \right]
\right. \nonumber\\
&-& 
\left. 
\beta^{(K,0),(K,0)}_{\;\;\;\;\downarrow\;\;\;\;\,,\;\;\;\uparrow}
S\left( \alpha^{(K,0)}_{\;\;\;\;\downarrow};T,K,q\right)
\right\}
\\
\frac{d}{dt} e^{(K,K),(K,K)}_{\;\;\;\;\uparrow\;\;\;,\;\;\;\;\downarrow} &=&
\frac{r}{K}c_{(K,K)} 
\left\{
-\beta^{(K,K),(K,K)}_{\;\;\;\;\uparrow\;\;\;\;\,,\;\;\;\;\downarrow}
S\left( \alpha^{(K,K)}_{\;\;\;\;\uparrow};T,K,q\right)
\right. \nonumber\\
&+& 
\left. 
\gamma^{(K,K),(K,K)}_{\;\;\;\;\uparrow\;\;\;\;\,,\;\;\;\;\uparrow}
\left[KR \left( \alpha^{(K,K)}_{\;\;\;\;\uparrow};T,q\right)
-S\left( \alpha^{(K,K)}_{\;\;\;\;\uparrow};T,K,q\right) \right]
\right\}
R \left( \alpha^{(K,K)}_{\;\;\;\;\uparrow};T,q\right) \nonumber\\
&+&
\frac{r}{K}  \left( 1-c_{(K,K)}\right)
\left\{
\gamma^{(K,K),(K,K)}_{\;\;\;\;\downarrow\;\;\;\;\,,\;\;\;\;\downarrow}
\left[KR \left( \alpha^{(K,K)}_{\;\;\;\;\downarrow};T,q\right)
-S\left( \alpha^{(K,K)}_{\;\;\;\;\downarrow};T,K,q\right) \right]
\right. \nonumber\\
&-& 
\left. 
\beta^{(K,K),(K,K)}_{\;\;\;\;\downarrow\;\;\;\;\,,\;\;\;\;\uparrow}
S\left( \alpha^{(K,K)}_{\;\;\;\;\downarrow};T,K,q\right)
\right\}
R \left( \alpha^{(K,K)}_{\;\;\;\;\downarrow};T,q\right)
\\
\frac{d}{dt} e^{(K,0),(K,K)}_{\;\;\;\;\uparrow\;\;\;,\;\;\;\;\uparrow} &=&
-\frac{1-r}{K}c_{(K,0)}
\left( 1-\gamma^{(K,0),(K,0)}_{\;\;\;\;\uparrow\;\;\;,\;\;\;\;\uparrow}\right)
\left[KR \left( \alpha^{(K,0)}_{\;\;\;\;\uparrow};T,q\right)
-S\left( \alpha^{(K,0)}_{\;\;\;\;\uparrow};T,K,q\right) \right]
\nonumber\\
&+&\frac{1-r}{K} \left( 1-c_{(K,0)}\right)
\left( 1-\beta^{(K,0),(K,0)}_{\;\;\;\;\downarrow\;\;\;,\;\;\;\;\uparrow} \right)
S\left( \alpha^{(K,0)}_{\;\;\;\;\downarrow};T,K,q\right)
\nonumber\\
&-&\frac{r}{K}c_{(K,K)} 
\left( 1-\gamma^{(K,K),(K,K)}_{\;\;\;\;\uparrow\;\;\;\;\,,\;\;\;\;\uparrow} \right)
\left[KR \left( \alpha^{(K,K)}_{\;\;\;\;\uparrow};T,q\right)
-S\left( \alpha^{(K,K)}_{\;\;\;\;\uparrow};T,K,q\right) \right]
R \left( \alpha^{(K,K)}_{\;\;\;\;\uparrow};T,q\right)
\nonumber\\
&+& \frac{r}{K}  \left( 1-c_{(K,K)}\right)
\left( 1-\beta^{(K,K),(K,K)}_{\;\;\;\;\downarrow\;\;\;\;\,,\;\;\;\;\uparrow}\right)
S\left( \alpha^{(K,K)}_{\;\;\;\;\downarrow};T,K,q\right)
R \left( \alpha^{(K,K)}_{\;\;\;\;\downarrow};T,q\right)
\\
\frac{d}{dt} e^{(K,0),(K,K)}_{\;\;\;\;\downarrow\;\;\;,\;\;\;\;\downarrow} &=&
\frac{1-r}{K}c_{(K,0)}
\left( 1-\beta^{(K,0),(K,0)}_{\;\;\;\;\uparrow\;\;\;,\;\;\;\;\downarrow}\right)
S\left( \alpha^{(K,0)}_{\;\;\;\;\uparrow};T,K,q\right) \nonumber\\
&-&\frac{1-r}{K} \left( 1-c_{(K,0)}\right)
\left( 1-\gamma^{(K,0),(K,0)}_{\;\;\;\;\downarrow\;\;\;,\;\;\;\;\downarrow} \right)
\left[KR \left( \alpha^{(K,0)}_{\;\;\;\;\downarrow};T,q\right)
-S\left( \alpha^{(K,0)}_{\;\;\;\;\downarrow};T,K,q\right) \right] \nonumber\\
&+&\frac{r}{K}c_{(K,K)} 
\left( 1-\beta^{(K,K),(K,K)}_{\;\;\;\;\uparrow\;\;\;\;\,,\;\;\;\;\downarrow} \right)
S\left( \alpha^{(K,K)}_{\;\;\;\;\uparrow};T,K,q\right)
R \left( \alpha^{(K,K)}_{\;\;\;\;\uparrow};T,q\right)
\nonumber\\
&-& \frac{r}{K}  \left( 1-c_{(K,K)}\right)
\left( 1-\gamma^{(K,K),(K,K)}_{\;\;\;\;\downarrow\;\;\;\;\,,\;\;\;\;\downarrow}\right)
\left[KR \left( \alpha^{(K,K)}_{\;\;\;\;\downarrow};T,q\right)
-S\left( \alpha^{(K,K)}_{\;\;\;\;\downarrow};T,K,q\right) \right]
R \left( \alpha^{(K,K)}_{\;\;\;\;\downarrow};T,q\right) \nonumber\\
\\
\frac{d}{dt} e^{(K,0),(K,K)}_{\;\;\;\;\uparrow\;\;\;,\;\;\;\;\downarrow} &=&
-\frac{1-r}{K}c_{(K,0)}
\left( 1-\beta^{(K,0),(K,0)}_{\;\;\;\;\uparrow\;\;\;,\;\;\;\;\downarrow}\right)
S\left( \alpha^{(K,0)}_{\;\;\;\;\uparrow};T,K,q\right) \nonumber\\
&+&\frac{1-r}{K} \left( 1-c_{(K,0)}\right)
\left( 1-\gamma^{(K,0),(K,0)}_{\;\;\;\;\downarrow\;\;\;,\;\;\;\;\downarrow} \right)
\left[KR \left( \alpha^{(K,0)}_{\;\;\;\;\downarrow};T,q\right)
-S\left( \alpha^{(K,0)}_{\;\;\;\;\downarrow};T,K,q\right) \right] \nonumber\\
&+&\frac{r}{K}c_{(K,K)} 
\left( 1-\gamma^{(K,K),(K,K)}_{\;\;\;\;\uparrow\;\;\;\;\,,\;\;\;\;\uparrow} \right)
\left[KR \left( \alpha^{(K,K)}_{\;\;\;\;\uparrow};T,q\right)
-S\left( \alpha^{(K,K)}_{\;\;\;\;\uparrow};T,K,q\right) \right]
R \left( \alpha^{(K,K)}_{\;\;\;\;\uparrow};T,q\right) \nonumber\\
&-& \frac{r}{K}  \left( 1-c_{(K,K)}\right)
\left( 1-\beta^{(K,K),(K,K)}_{\;\;\;\;\downarrow\;\;\;\;\,,\;\;\;\;\uparrow}\right)
S\left( \alpha^{(K,K)}_{\;\;\;\;\downarrow};T,K,q\right)
R \left( \alpha^{(K,K)}_{\;\;\;\;\downarrow};T,q\right)
\\
\frac{d}{dt} e^{(K,0),(K,K)}_{\;\;\;\;\downarrow\;\;\;,\;\;\;\;\uparrow} &=&
\frac{1-r}{K}c_{(K,0)}
\left( 1-\gamma^{(K,0),(K,0)}_{\;\;\;\;\uparrow\;\;\;,\;\;\;\;\uparrow}\right)
\left[KR \left( \alpha^{(K,0)}_{\;\;\;\;\uparrow};T,q\right)
-S\left( \alpha^{(K,0)}_{\;\;\;\;\uparrow};T,K,q\right) \right]
\nonumber\\
&-&\frac{1-r}{K} \left( 1-c_{(K,0)}\right)
\left( 1-\beta^{(K,0),(K,0)}_{\;\;\;\;\downarrow\;\;\;,\;\;\;\;\uparrow} \right)
S\left( \alpha^{(K,0)}_{\;\;\;\;\downarrow};T,K,q\right)
\nonumber\\
&-&\frac{r}{K}c_{(K,K)} 
\left( 1-\beta^{(K,K),(K,K)}_{\;\;\;\;\uparrow\;\;\;\;\,,\;\;\;\;\downarrow} \right)
S\left( \alpha^{(K,K)}_{\;\;\;\;\uparrow};T,K,q\right)
R \left( \alpha^{(K,K)}_{\;\;\;\;\uparrow};T,q\right)
\nonumber\\
&+& \frac{r}{K}  \left( 1-c_{(K,K)}\right)
\left( 1-\gamma^{(K,K),(K,K)}_{\;\;\;\;\downarrow\;\;\;\;\,,\;\;\;\;\downarrow}\right)
\left[KR \left( \alpha^{(K,K)}_{\;\;\;\;\downarrow};T,q\right)
-S\left( \alpha^{(K,K)}_{\;\;\;\;\downarrow};T,K,q\right) \right]
R \left( \alpha^{(K,K)}_{\;\;\;\;\downarrow};T,q\right),
\nonumber\\
\label{dek0kkdt_het}
\end{eqnarray}
where the significant conditional probabilities are 
\begin{eqnarray}
    \alpha^{(K,0)}_{\;\;\;\;\downarrow} &=&
    \frac{e^{(K,0),(K,0)}_{\;\;\;\;\downarrow\;\;\;,\;\;\;\;\uparrow}+
    e^{(K,0),(K,K)}_{\;\;\;\;\downarrow\;\;\;,\;\;\;\;\uparrow}}
    {e^{(K,0),(K,0)}_{\;\;\;\;\downarrow\;\;\;,\;\;\;\;\uparrow}+
    e^{(K,0),(K,K)}_{\;\;\;\;\downarrow\;\;\;,\;\;\;\;\uparrow}+
    e^{(K,0),(K,0)}_{\;\;\;\;\downarrow\;\;\;,\;\;\;\;\downarrow}+
    e^{(K,0),(K,K)}_{\;\;\;\;\downarrow\;\;\;,\;\;\;\;\downarrow}} 
    \nonumber\\
    \alpha^{(K,0)}_{\;\;\;\;\uparrow} &=&
    \frac{e^{(K,0),(K,0)}_{\;\;\;\;\uparrow\;\;\;,\;\;\;\;\downarrow}+
    e^{(K,0),(K,K)}_{\;\;\;\;\uparrow\;\;\;,\;\;\;\;\downarrow}}
    {e^{(K,0),(K,0)}_{\;\;\;\;\uparrow\;\;\;,\;\;\;\;\downarrow}+
    e^{(K,0),(K,K)}_{\;\;\;\;\uparrow\;\;\;,\;\;\;\;\downarrow}+
    e^{(K,0),(K,0)}_{\;\;\;\;\uparrow\;\;\;,\;\;\;\;\uparrow}+
    e^{(K,0),(K,K)}_{\;\;\;\;\uparrow\;\;\;,\;\;\;\;\uparrow}} 
    \nonumber\\
    \alpha^{(K,K)}_{\;\;\;\;\downarrow} &=&
    \frac{e^{(K,K),(K,K)}_{\;\;\;\;\downarrow\;\;\;,\;\;\;\;\uparrow}+
    e^{(K,K),(K,0)}_{\;\;\;\;\downarrow\;\;\;,\;\;\;\;\uparrow}}
    {e^{(K,K),(K,K)}_{\;\;\;\;\downarrow\;\;\;,\;\;\;\;\uparrow}+
    e^{(K,K),(K,0)}_{\;\;\;\;\downarrow\;\;\;,\;\;\;\;\uparrow}+
    e^{(K,K),(K,K)}_{\;\;\;\;\downarrow\;\;\;,\;\;\;\;\downarrow}+
    e^{(K,K),(K,0)}_{\;\;\;\;\downarrow\;\;\;,\;\;\;\;\downarrow}} 
    \nonumber\\
    \alpha^{(K,K)}_{\;\;\;\;\uparrow} &=&
    \frac{e^{(K,K),(K,K)}_{\;\;\;\;\uparrow\;\;\;,\;\;\;\;\downarrow}+
    e^{(K,K),(K,0)}_{\;\;\;\;\uparrow\;\;\;,\;\;\;\;\downarrow}}
    {e^{(K,K),(K,K)}_{\;\;\;\;\uparrow\;\;\;,\;\;\;\;\downarrow}+
    e^{(K,K),(K,0)}_{\;\;\;\;\uparrow\;\;\;,\;\;\;\;\downarrow}+
    e^{(K,K),(K,K)}_{\;\;\;\;\uparrow\;\;\;,\;\;\;\;\uparrow}+
    e^{(K,K),(K,0)}_{\;\;\;\;\uparrow\;\;\;,\;\;\;\;\uparrow}},
\end{eqnarray}
\begin{eqnarray}
    \beta^{(K,0),(K,0)}_{\;\;\;\;\downarrow\;\;\;\;\,,\;\;\;\;\uparrow} =
    \frac{e^{(K,0),(K,0)}_{\;\;\;\;\downarrow\;\;\;,\;\;\;\;\uparrow}}
    {e^{(K,0),(K,0)}_{\;\;\;\;\downarrow\;\;\;,\;\;\;\;\uparrow}+
    e^{(K,0),(K,K)}_{\;\;\;\;\downarrow\;\;\;,\;\;\;\;\uparrow}} 
    &,&
    \beta^{(K,0),(K,0)}_{\;\;\;\;\uparrow\;\;\;\;\,,\;\;\;\;\downarrow} =
    \frac{e^{(K,0),(K,0)}_{\;\;\;\;\uparrow\;\;\;,\;\;\;\;\downarrow}}
    {e^{(K,0),(K,0)}_{\;\;\;\;\uparrow\;\;\;,\;\;\;\;\downarrow}+
    e^{(K,0),(K,K)}_{\;\;\;\;\uparrow\;\;\;,\;\;\;\;\downarrow}}
    \nonumber\\
    \beta^{(K,K),(K,K)}_{\;\;\;\;\downarrow\;\;\;\;\,,\;\;\;\;\uparrow} =
    \frac{e^{(K,K),(K,K)}_{\;\;\;\;\downarrow\;\;\;,\;\;\;\;\uparrow}}
    {e^{(K,K),(K,K)}_{\;\;\;\;\downarrow\;\;\;,\;\;\;\;\uparrow}+
    e^{(K,K),(K,0)}_{\;\;\;\;\downarrow\;\;\;,\;\;\;\;\uparrow}} 
    &,&
    \beta^{(K,K),(K,K)}_{\;\;\;\;\uparrow\;\;\;\;\,,\;\;\;\;\downarrow} =
    \frac{e^{(K,K),(K,K)}_{\;\;\;\;\uparrow\;\;\;,\;\;\;\;\downarrow}}
    {e^{(K,K),(K,K)}_{\;\;\;\;\uparrow\;\;\;,\;\;\;\;\downarrow}+
    e^{(K,K),(K,0)}_{\;\;\;\;\uparrow\;\;\;,\;\;\;\;\downarrow}},
\end{eqnarray}
\begin{eqnarray}
    \gamma^{(K,0),(K,0)}_{\;\;\;\;\downarrow\;\;\;\;\,,\;\;\;\;\downarrow} =
    \frac{e^{(K,0),(K,0)}_{\;\;\;\;\downarrow\;\;\;,\;\;\;\;\downarrow}}
    {e^{(K,0),(K,0)}_{\;\;\;\;\downarrow\;\;\;,\;\;\;\;\downarrow}+
    e^{(K,0),(K,K)}_{\;\;\;\;\downarrow\;\;\;,\;\;\;\;\downarrow}} 
    &,&
    \gamma^{(K,0),(K,0)}_{\;\;\;\;\uparrow\;\;\;\;\,,\;\;\;\;\uparrow} =
    \frac{e^{(K,0),(K,0)}_{\;\;\;\;\uparrow\;\;\;,\;\;\;\;\uparrow}}
    {e^{(K,0),(K,0)}_{\;\;\;\;\uparrow\;\;\;,\;\;\;\;\uparrow}+
    e^{(K,0),(K,K)}_{\;\;\;\;\uparrow\;\;\;,\;\;\;\;\uparrow}}
    \nonumber\\
    \gamma^{(K,K),(K,K)}_{\;\;\;\;\downarrow\;\;\;\;\,,\;\;\;\;\downarrow} =
    \frac{e^{(K,K),(K,K)}_{\;\;\;\;\downarrow\;\;\;,\;\;\;\;\downarrow}}
    {e^{(K,K),(K,K)}_{\;\;\;\;\downarrow\;\;\;,\;\;\;\;\downarrow}+
    e^{(K,K),(K,0)}_{\;\;\;\;\downarrow\;\;\;,\;\;\;\;\downarrow}} 
    &,&
    \gamma^{(K,K),(K,K)}_{\;\;\;\;\uparrow\;\;\;\;\,,\;\;\;\;\uparrow} =
    \frac{e^{(K,K),(K,K)}_{\;\;\;\;\uparrow\;\;\;,\;\;\;\;\uparrow}}
    {e^{(K,K),(K,K)}_{\;\;\;\;\uparrow\;\;\;,\;\;\;\;\uparrow}+
    e^{(K,K),(K,0)}_{\;\;\;\;\uparrow\;\;\;,\;\;\;\;\uparrow}}.
\end{eqnarray}
Concentration $\tilde{c}$ of spins directed up within each layer and concentration $c$ of spins directed up in the MN are defined in the same way as in Sec.\ \ref{HoPA}. Natural initial conditions for the system of equations (\ref{dcK0dt_het} - \ref{dek0kkdt_het}) are
$c_{(K,0)}(0) = c_{(K,K)}(0)= \rho_0$,
$e^{(K,0),(K,0)}_{\;\;\;\;\uparrow\;\;\;,\;\;\;\;\uparrow}(0)= (1-r)^2 \rho_{0}^{2}$,
$e^{(K,0),(K,0)}_{\;\;\;\;\uparrow\;\;\;,\;\;\;\;\downarrow}(0)= (1-r)^2 \rho_{0} (1-\rho_0)$,
$e^{(K,0),(K,0)}_{\;\;\;\;\downarrow\;\;\;,\;\;\;\;\downarrow}(0)= (1-r)^2 (1-\rho_0)^2$,
$e^{(K,K),(K,K)}_{\;\;\;\;\uparrow\;\;\;,\;\;\;\;\uparrow}(0)= r^2 \rho_{0}^{2}$,
$e^{(K,K),(K,K)}_{\;\;\;\;\uparrow\;\;\;,\;\;\;\;\downarrow}(0)= r^2 \rho_{0} (1-\rho_0)$,
$e^{(K,K),(K,K)}_{\;\;\;\;\downarrow\;\;\;,\;\;\;\;\downarrow}(0)= r^2 (1-\rho_0)^2$,
$e^{(K,0),(K,K)}_{\;\;\;\;\uparrow\;\;\;,\;\;\;\;\uparrow}(0)= (1-r)r \rho_{0}^{2}$,
$e^{(K,0),(K,K)}_{\;\;\;\;\downarrow\;\;\;,\;\;\;\;\downarrow}(0)= (1-r)r (1-\rho_0)^2$,
$e^{(K,0),(K,K)}_{\;\;\;\;\uparrow\;\;\;,\;\;\;\;\downarrow}(0)= (1-r)^2 \rho_{0} (1-\rho_0)=
e^{(K,0),(K,0)}_{\;\;\;\;\downarrow\;\;\;,\;\;\;\;\uparrow}(0)= (1-r)r \rho_{0} (1-\rho_0)$,
where $\rho_0$ can be chosen arbitrarily.

\begin{figure}
    \centering
    \includegraphics[width=0.5\linewidth]{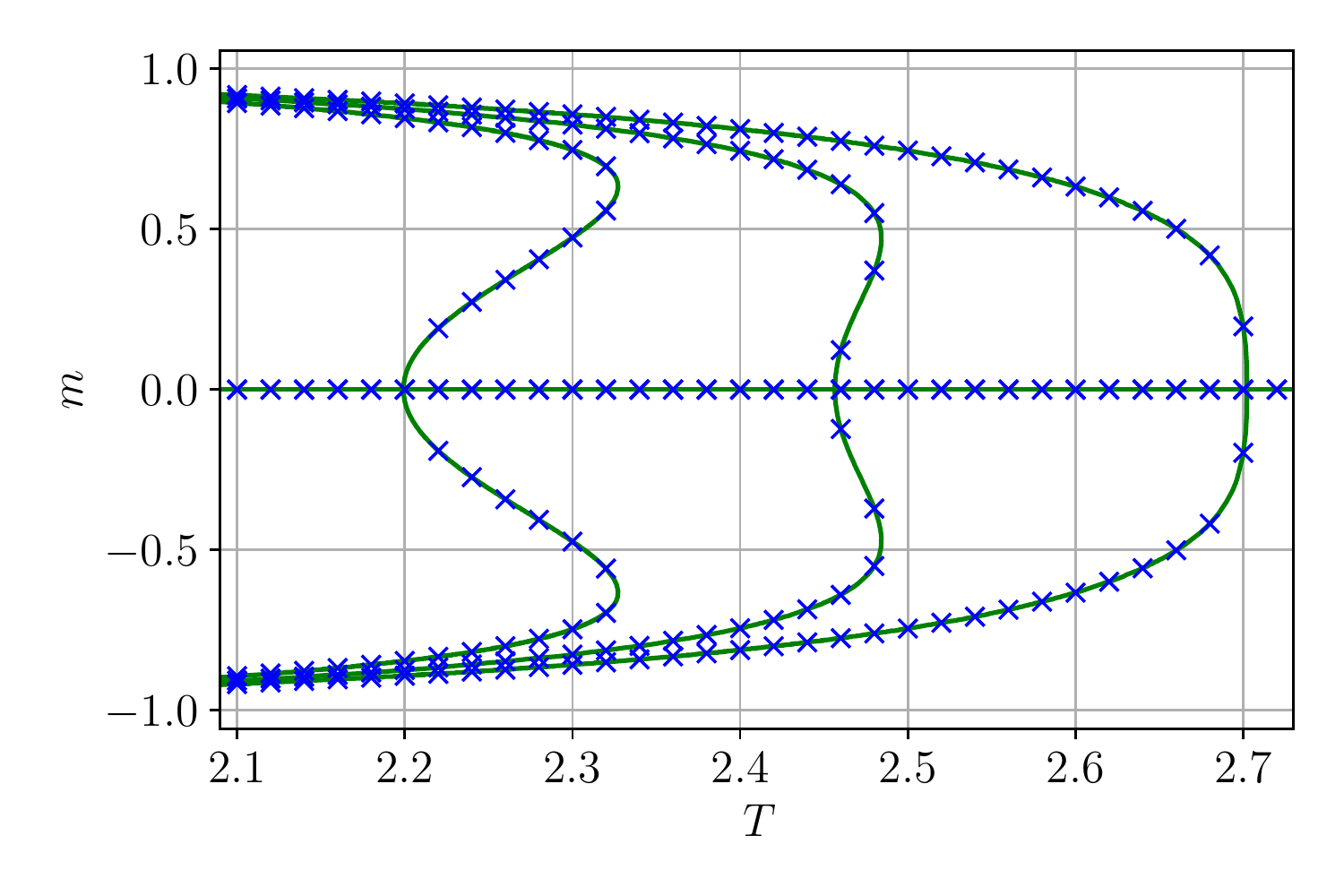}
    \caption{Magnetization $m$ vs.\ temperature $T$ obtained from the homogeneous PA (solid lines) and from the heterogeneous PA (symbols) for $q=4$, $K=20$ and $r=0.1,0.15,0.2$ (from left to right).}
    \label{Fig:HoPAHetPA}
\end{figure}

As mentioned in Sec.\ \ref{HoPA} the magnetization curves obtained from the fully heterogeneous PA are practically indistinguishable from those obtained from the homogeneous PA. This is illustrated by examples in Fig.\ \ref{Fig:HoPAHetPA}. 
A question arises whether the two above-mentioned kinds of the PA will yield almost indistinguishable results also in the case of the $q$-neighbor Ising model on MNs with the more general multidegree distribution (\ref{JointPDgen}), in particular when the layers are heterogeneous (e.g, scale-free) networks. In this context let us mention that in the related case of the voter model on scale-free networks heterogeneous PA yields predictions noticeably different from those of the homogeneous PA \cite{Pugliese09}. Unfortunately, in the case of models on MNs verifying this difference requires integration of prohibitively large systems of equations (\ref{rate1hetero}), (\ref{rateeehetero}), (\ref{ratee-ehetero}) which is beyond the scope of this work. Alternatively, as in Ref.\ \cite{Jedrzejewski22}, one can consider only two classes of nodes, those belonging or not to the overlap, and perform averaging over the multidegree distribution in the equations for the corresponding densities of bonds. Then, as discussed in Sec.\ \ref{HoPA} for the case of the homogeneous PA, the equations of motion for the macroscopic quantities will have the form of Eq.\ (\ref{dcK0dt_het}-\ref{dek0kkdt_het}) with $K$ replaced by the mean degree of nodes within layers $\tilde{\langle k \rangle}$. Hence, predictions of such heterogeneous PA again will not depend on the specific form of the multidegree distribution and remain similar to those of the homogeneous PA. 

\subsection{AMEs-based heterogeneous pair approximation}

\label{appendixB}

The AMEs-based heterogeneous PA again uses the assumption that the probability that a spin directed up or down in a node with multidegree $\mathbf{k}$ has a given number of neighboring spins directed up obeys a binomial distribution; for models on MNs this assumption is made for each layer separately, and the related binomial distributions are assumed to be independent. Hence, in contrast with the homogeneous PA, in the AMEs-based heterogeneous PA it is taken into account that for a node with multidegree $\mathbf{k}$ occupied by a spin with downward or upward direction the respective probabilities $\vartheta^{(L)}_{\mathbf{k}}$, $\eta^{(L)}_{\mathbf{k}}$ that a randomly chosen neighboring node within the layer $G^{(L)}$ is occupied by a spin directed upward can depend on $\mathbf{k}$. However, in contrast with the fully heterogeneous PA developed in Appendix A, all active or inactive edges attached to a given node within a given layer are treated in the same way and obey common binomial distributions \cite{Gleeson11,Gleeson13}. As mentioned in Sec.\ \ref{AMEs}, these two assumptions should make the AMEs-based heterogeneous PA more accurate than the homogeneous PA and less accurate than the fully heterogeneous PA. Eventually, in the AMEs-based heterogeneous PA the time-dependent macroscopic quantities are the density 
$c_{\mathbf{k}}$ of spins directed up in nodes with multidegree $\mathbf{k}$ as well as the above-mentioned probabilities $\vartheta^{(L)}_{\mathbf{k}}$, $\eta^{(L)}_{\mathbf{k}}$. 

In terms of the densities $c_{\mathbf{k,m}}$ and  $s_{\mathbf{k,m}}$ used in the AMEs, Eq.\ (\ref{dskmdt}), (\ref{dikmdt}) the above-mentioned macroscopic quantities can be expressed as $c_{\mathbf{k}}= \sum_{\mathbf{m}} c_{\mathbf{k,m}} = 1-\sum_{\mathbf{m}} s_{\mathbf{k,m}}$, $\vartheta^{(L)}_{\mathbf{k}}=\sum_{\mathbf{m}}m^{(L)} s_{\mathbf{k,m}}/\left[ k^{(L)} \left( 1- c_{\mathbf{k}}\right)\right]$,
$\eta^{(L)}_{\mathbf{k}}=\sum_{\mathbf{m}}m^{(L)} c_{\mathbf{k,m}}/\left[ k^{(L)} c_{\mathbf{k}}\right]$.
Then, the core approximation for the AMEs-based heterogeneous PA can be made, according to which 
$s_{\mathbf{k,m}} \approx \left( 1- c_{\mathbf{k}}\right)
\prod_{L=A}^{L_{\max}}
B_{k^{(L)},m^{(L)}}\left( \vartheta_{\mathbf{k}}^{(L)} \right)$,
$c_{\mathbf{k,m}} \approx c_{\mathbf{k}}
\prod_{L=A}^{L_{\max}}
B_{k^{(L)},m^{(L)}}\left( \eta_{\mathbf{k}}^{(L)} \right)$.
The latter approximation should be made in  Eq.\ (\ref{dskmdt}), (\ref{dikmdt}) as well as in the definitions of the average rates $\beta_{s}^{(L)}, \ldots, \gamma_{c}^{(L)}$, so that, e.g.,
$\beta_{s}^{(L')} \approx \bar{\beta}_{s}^{(L')} =
\big \langle\left( 1- c_{\mathbf{k}}\right) \sum_{\mathbf{m}}\left( k^{(L')}-m^{(L')} \right) F_{\mathbf{k,m}} 
\prod_{L=A}^{L_{\max}}
B_{k^{(L)},m^{(L)}}\left( \vartheta_{\mathbf{k}}^{(L)} \right)
\big \rangle/
\big \langle \left( 1- c_{\mathbf{k}}\right)k^{(L')}\left( 1-\vartheta_{\mathbf{k}}^{(L')} \right)\big \rangle$, etc. Differentiating the definitions of $c_{\mathbf{k}}$, $\vartheta^{(L)}_{\mathbf{k}}$, $\eta^{(L)}_{\mathbf{k}}$ with respect to time and using  Eq.\ (\ref{dskmdt}), (\ref{dikmdt}) with the above-mentioned approximations yields the following 
system of equations for the time dependence of the macroscopic quantities in the heterogeneous PA,
\begin{eqnarray}
\frac{d c_{\mathbf{k}}}{dt} &=& 
- c_{\mathbf{k}}\sum_{\mathbf{m}}R_{\mathbf{k,m}} 
\prod_{L=A}^{L_{\max}}
B_{k^{(L)},m^{(L)}}\left( \eta_{\mathbf{k}}^{(L)} \right) 
+ \left( 1 - c_{\mathbf{k}}\right) \sum_{\mathbf{m}}F_{\mathbf{k,m}} 
\prod_{L=A}^{L_{\max}}
B_{k^{(L)},m^{(L)}}\left( \vartheta_{\mathbf{k}}^{(L)} \right) ,\;\;\;\;\;
\label{dckdt}\\
\frac{d \vartheta_{\mathbf{k}}^{(L')}}{dt} &=& 
\sum_{\mathbf{m}}\left( \vartheta_{\mathbf{k}}^{(L')}-\frac{m^{(L')}}{k^{(L')}}
\right)
\left[ 
F_{\mathbf{k,m}} 
\prod_{L=A}^{L_{\max}}
B_{k^{(L)},m^{(L)}}\left( \vartheta_{\mathbf{k}}^{(L)}\right) - \frac{c_{\mathbf{k}}}{1-c_{\mathbf{k}}} 
R_{\mathbf{k,m}} 
\prod_{L=A}^{L_{\max}}
B_{k^{(L)},m^{(L)}}\left( \eta_{\mathbf{k}}^{(L)} \right)
\right] \nonumber\\
&&+\bar{\beta}_{s}^{(L')}
\left( 1-\vartheta_{\mathbf{k}}^{(L')} \right) -\bar{\gamma}_{s}^{(L')} 
\vartheta_{\mathbf{k}}^{(L')},
\label{dpkdt}
\\
\frac{d \eta_{\mathbf{k}}^{(L')}}{dt} &=& 
\sum_{\mathbf{m}}\left( \eta_{\mathbf{k}}^{(L')}-\frac{m^{(L')}}{k^{(L')}}\right)
\left[ 
R_{\mathbf{k,m}} 
\prod_{L=A}^{L_{\max}}
B_{k^{(L)},m^{(L)}}\left( \eta_{\mathbf{k}}^{(L)} \right)
- \frac{1- c_{\mathbf{k}}}{c_{\mathbf{k}}} 
F_{\mathbf{k,m}} 
\prod_{L=A}^{L_{\max}}
B_{k^{(L)},m^{(L)}}\left( \vartheta_{\mathbf{k}}^{(L)}\right)
\right] \nonumber\\
&& +\bar{\beta}_{i}^{(L')}
\left( 1-\eta_{\mathbf{k}}^{(L')} \right) -\bar{\gamma}_{i}^{(L')} 
\eta_{\mathbf{k}}^{(L')},
\label{dqkdt}
\end{eqnarray}
where $L'=A,B\ldots L_{\max}$. The above equations are very similar to those obtained in the AMEs-based heterogeneous PA for the spin models on (monoplex) networks \cite{Gleeson11,Gleeson13}; in particular, terms containing $\beta_{s}^{(L)},\ldots, \gamma_{i}^{(L)}$ with $L\neq L'$ do not occur in Eq.\ (\ref{dpkdt}), (\ref{dqkdt}) for $\vartheta_{\mathbf{k}}^{(L')}$, $\eta_{\mathbf{k}}^{(L')}$ since the respective terms from  Eq.\ (\ref{dskmdt}), (\ref{dikmdt}) sum up to zero in the derivation. It should be mentioned that the AMEs can also be a starting point to obtain the homogeneous PA from Sec.\ \ref{HoPA} by assuming that the probability that a spin directed down has within the layer $G^{(L)}$ a neighboring spin directed up does not depend on $\mathbf{k}$ and can be expressed as the average $\theta_{\downarrow}^{(L)} = \langle \sum_{\mathbf{m}}m^{(L)} s_{\mathbf{k,m}}\rangle/\langle k^{(L)} \left( 1- c_{\mathbf{k}}\right)\rangle$ \cite{Gleeson11,Gleeson13}. 

In the case of the $q$-neighbor Ising model on MNs with partial overlap of nodes and with layers in the form of RRGs, with the multidegree distribution $P\left( \mathbf{k}\right)$ given by Eq.\ (\ref{JointPD}), there are three classes of nodes with $\mathbf{k}=(K,0)$, $\mathbf{k}=(0,K)$ and $\mathbf{k}=(K,K)$, and two layers $G^{(L)}$, $L=A,B$, thus the system of equations (\ref{dckdt}-\ref{dqkdt}) is 11-dimensional. Due to the symmetry of the model solutions of these equations should be constrained to a subspace $c_{(0,K)}=c_{(K,0)}$, $\vartheta_{(K,0)}^{(A)}=\vartheta_{(0,K)}^{(B)}\equiv \vartheta_{(0,K)}$, $\eta_{(K,0)}^{(A)}=\eta_{(0,K)}^{(B)}\equiv \eta_{(0,K)}$, $\vartheta_{(K,K)}^{(A)}=\vartheta_{(K,K)}^{(B)}\equiv \vartheta_{(K,K)}$, $\eta_{(K,K)}^{(A)}=\eta_{(K,K)}^{(B)}\equiv \eta_{(K,K)}$ which reduces the number of equations to six. Performing summations in Eq.\ (\ref{dckdt}-\ref{dqkdt}) as in Ref.\ \cite{Chmiel18} the following system of equations for the macroscopic quantities is obtained in the AMEs-based heterogeneous PA for the model under study,
\begin{eqnarray}
\frac{d c_{(K,0)}}{dt} &=& 
-c_{(K,0)}R\left( 1-\eta_{(K,0)};T,q\right)  
+\left( 1-c_{(K,0)}\right)R \left( \vartheta_{(K,0)};T,q\right), 
\label{dck0dt}\\
\frac{d \vartheta_{(K,0)}}{dt} &=& 
\vartheta_{(K,0)}\left[R\left( \vartheta_{(K,0)};T,q\right) -
\frac{c_{(K,0)}}{1-c_{(K,0)}} R\left( 1-\eta_{(K,0)};T,q\right) \right] \nonumber\\
&-& \frac{1}{K}\left\{S\left( \vartheta_{(K,0)};T,K,q\right) 
-\frac{c_{(K,0)}}{1-c_{(K,0)}} 
\left[ KR\left( 1-\eta_{(K,0)};T,q\right) 
-S\left( 1-\eta_{(K,0)};T,K,q\right) \right]\right\}
\nonumber\\
&+&\bar{\beta}_{s}\left( 1-\vartheta_{(K,0)}\right) -\bar{\gamma}_{s}\vartheta_{(K,0)},
\\
\frac{d \eta_{(K,0)}}{dt} &=& 
\eta_{(K,0)}\left[ R\left( 1-\eta_{(K,0)};T,q\right) -
\frac{1- c_{(K,0)}}{c_{(K,0)}} R\left( \vartheta_{(K,0)};T,q\right)  \right] \nonumber\\
&-& \frac{1}{K}\left\{\left[KR\left( 1-\eta_{(K,0)};T,q\right) - S\left( 1-\eta_{(K,0)};T,K,q\right)\right]
-\frac{1- c_{(K,0)}}{c_{(K,0)}} 
S\left( \vartheta_{(K,0)};T,K,q\right) \right\}
\nonumber\\
&+&\bar{\beta}_{i}\left( 1-\eta_{(K,0)}\right) -\bar{\gamma}_{i}\eta_{(K,0)},
\\
\frac{d c_{(K,K)}}{dt} &=& 
-c_{(K,K)}\left[ R\left( 1-\eta_{(K,K)};T,q\right) \right]^2
+\left( 1-c_{(K,K)}\right)\left[R\left( \vartheta_{(K,K)};T,q\right)\right]^2,
\\
\frac{d \vartheta_{(K,K)}}{dt} &=& 
\vartheta_{(K,K)}\left\{\left[R\left( \vartheta_{(K,K)};T,q\right)  \right]^2-
\frac{c_{(K,K)}}{1-c_{(K,K)}} \left[R\left( 1-\eta_{(K,K)};T,q\right)\right]^2 \right\} \nonumber\\
&-& \frac{1}{K}\left\{S\left( \vartheta_{(K,K)};T,K,q\right) 
R\left( \vartheta_{(K,K)};T,q\right)
\right. \nonumber\\
&& \left. \;\;\;\;\; -\frac{c_{(K,K)}}{1-c_{(K,K)}} 
\left[ KR\left( 1-\eta_{(K,0)};T,q\right) 
-S\left( 1-\eta_{(K,0)};T,K,q\right)\right]    
R\left( 1-\eta_{(K,K)};T,q\right)
\right\} \nonumber\\
&+&\bar{\beta}_{s}\left( 1-\vartheta_{(K,K)}\right) -\bar{\gamma}_{s}\vartheta_{(K,K)},
\\
\frac{d \eta_{(K,K)}}{dt} &=& 
\eta_{(K,K)}\left\{\left[R\left( 1-\eta_{(K,K)};T,q\right) \right]^2-
\frac{1- c_{(K,K)}}{c_{(K,K)}}  \left[R\left( \vartheta_{(K,K)};T,q\right) \right]^2 \right\} \nonumber\\
&-& \frac{1}{K}\left\{\left[
KR\left( 1-\eta_{(K,0)};T,q\right) - 
S\left( 1-\eta_{(K,0)};T,K,q\right)\right]
R\left( 1-\eta_{(K,K)};T,q\right) 
\right. \nonumber\\
&& \left. \;\;\;\;\; -\frac{1- c_{(K,K)}}{c_{(K,K)}}
S\left( \vartheta_{(K,K)};T,K,q\right) 
R\left( \vartheta_{(K,K)};T,q\right) 
\right\}
+\bar{\beta}_{i}\left( 1-\eta_{(K,K)}\right) -\bar{\gamma}_{i}\eta_{(K,K)},
\label{dqkkdt}
\end{eqnarray}
where the average rates are
\begin{eqnarray}
\bar{\beta}_{s}&=&
\left[\frac{1-r}{2-r}\left(1-c_{(K,0)}\right)K\left( 1-\vartheta_{(K,0)}\right)
+\frac{r}{2-r}\left(1-c_{(K,K)}\right)K\left( 1-\vartheta_{(K,K)}\right) \right]^{-1}
\nonumber\\
&\times& 
\left\{
\frac{1-r}{2-r}\left(1-c_{(K,0)}\right)
\left[ KR\left( \vartheta_{(K,0)};T,q\right) -
S\left( \vartheta_{(K,0)};T,K,q\right)\right]
\right. \nonumber\\
&&\left.
+ \frac{r}{2-r}\left(1-c_{(K,K)}\right) 
\left[ KR\left( \vartheta_{(K,K)};T,q\right) 
-S\left( \vartheta_{(K,K)};T,K,,q\right) \right]
R \left( \vartheta_{(K,K)};T,q\right)\right\},
\\
\bar{\gamma}_{s}&=&
\left[\frac{1-r}{2-r}c_{(K,0)}K\left( 1-\eta_{(K,0)}\right)
+\frac{r}{2-r}c_{(K,K)}K\left( 1-\eta_{(K,K)}\right) \right]^{-1}
\nonumber\\
&\times& 
\left\{
\frac{1-r}{2-r}c_{(K,0)}
S \left( 1-\eta_{(K,0)};T,K,q\right)
+ \frac{r}{2-r}c_{(K,K)} 
S \left( 1-\eta_{(K,K)};T,K,q\right) 
R\left( 1-\eta_{(K,K)};T,q\right)\right\},
\\
\bar{\beta}_{i}&=&
\left[\frac{1-r}{2-r}\left(1-c_{(K,0)}\right)K \vartheta_{(K,0)}
+\frac{r}{2-r}\left(1-c_{(K,K)}\right)K \vartheta_{(K,K)}\right]^{-1}
\nonumber\\
&\times& 
\left\{
\frac{1-r}{2-r}\left(1-c_{(K,0)}\right)
S \left( \vartheta_{(K,0)};T,K,q\right) 
+ \frac{r}{2-r}\left(1-c_{(K,K)}\right) 
S \left( \vartheta_{(K,K)};T,K,q\right)
R\left( \vartheta_{(K,K)};T,q\right)\right\},
\\
\bar{\gamma}_{i}&=&
\left[\frac{1-r}{2-r}c_{(K,0)}K\eta_{(K,0)}
+\frac{r}{2-r}c_{(K,K)}K\eta_{(K,K)}\right]^{-1}
\nonumber\\
&\times& 
\left\{
\frac{1-r}{2-r}c_{(K,0)}
\left[ KR \left( 1-\eta_{(K,0)};T,q\right) - 
S \left( 1-\eta_{(K,0)};T,K,q\right) \right]
\right. \nonumber\\
&&\left.
+ \frac{r}{2-r}c_{(K,K)} 
\left[ KR \left( 1-\eta_{(K,K)};T,q\right) - 
S \left( 1-\eta_{(K,K)};T,K,q\right) \right]
R \left( 1-\eta_{(K,K)};T,q\right)\right\}.\nonumber\\
&&
\end{eqnarray}
Concentration $\tilde{c}$ of spins directed up within each layer and concentration $c$ of spins directed up in the MN are defined in the same way as in Sec.\ \ref{HoPA}. Natural initial conditions for the system of equations (\ref{dck0dt}-\ref{dqkkdt}) are $\vartheta_{(K,0)}(0)=\eta_{(K,0)}(0)=
\vartheta_{(K,K)}(0)=\eta_{(K,K)}(0)=\tilde{c}(0)$, while $c_{(K,0)}(0)$, $c_{(K,K)}(0)$ can be chosen arbitrarily. 

% BBBBBBBBBBBBBBBBBBBBBBBBBBBBBBBBBBBBBBBBBBBBBBBBBBBBBBB%

\end{document}